\documentclass[twocolumn]{aastex701}

\usepackage[perpage]{footmisc}

\submitjournal{ApJ}

\usepackage{savesym}
\savesymbol{leftroot}
\savesymbol{uproot}
\savesymbol{iint}
\savesymbol{iiint}
\savesymbol{iiiint}
\savesymbol{idotsint}
\usepackage{amsmath}
\restoresymbol{AMS}{leftroot}
\restoresymbol{AMS}{uproot}
\restoresymbol{AMS}{iint}
\restoresymbol{AMS}{iiint}
\restoresymbol{AMS}{iiiint}
\restoresymbol{AMS}{idotsint}
\usepackage{soul}

\usepackage{xurl}
\begin{document} 

\title{ 
A XRISM/Resolve view of the dynamics in the hot gaseous atmosphere of M87 }

\suppressAffiliations
\correspondingauthor{Hannah McCall, Aurora Simionescu, Caroline Kilbourne\\
\hspace*{-4mm}Authors' affilliations are given at the end of the preprint.}

\collaboration{0}{XRISM Collaboration:%
}

\author[0000-0003-4721-034X]{Marc Audard}
\affiliation{Department of Astronomy, University of Geneva, Versoix CH-1290, Switzerland} 
\email{Marc.Audard@unige.ch}

\author{Hisamitsu Awaki}
\affiliation{Department of Physics, Ehime University, Ehime 790-8577, Japan}
\email{awaki@astro.phys.sci.ehime-u.ac.jp}

\author[0000-0002-1118-8470]{Ralf Ballhausen}
\affiliation{Department of Astronomy, University of Maryland, College Park, MD 20742, USA}
\affiliation{NASA / Goddard Space Flight Center, Greenbelt, MD 20771, USA}
\affiliation{Center for Research and Exploration in Space Science and Technology, NASA / GSFC (CRESST II), Greenbelt, MD 20771, USA}
\email{ballhaus@umd.edu}

\author[0000-0003-0890-4920]{Aya Bamba}
\affiliation{Department of Physics, University of Tokyo, Tokyo 113-0033, Japan}
\email{bamba@phys.s.u-tokyo.ac.jp}

\author[0000-0001-9735-4873]{Ehud Behar}
\affiliation{Department of Physics, Technion, Technion City, Haifa 3200003, Israel}
\email{behar@physics.technion.ac.il}

\author[0000-0003-2704-599X]{Rozenn Boissay-Malaquin}
\affiliation{Center for Space Sciences and Technology, University of Maryland, Baltimore County (UMBC), Baltimore, MD, 21250 USA}
\affiliation{NASA / Goddard Space Flight Center, Greenbelt, MD 20771, USA}
\affiliation{Center for Research and Exploration in Space Science and Technology, NASA / GSFC (CRESST II), Greenbelt, MD 20771, USA}
\email{rozennbm@umbc.edu}

\author[0000-0003-2663-1954]{Laura Brenneman}
\affiliation{Center for Astrophysics | Harvard-Smithsonian, Cambridge, MA 02138, USA}
\email{lbrenneman@cfa.harvard.edu}

\author[0000-0001-6338-9445]{Gregory V.\ Brown}
\affiliation{Lawrence Livermore National Laboratory, Livermore, CA 94550, USA}
\email{brown86@llnl.gov}

\author[0000-0002-5466-3817]{Lia Corrales}
\affiliation{Department of Astronomy, University of Michigan, Ann Arbor, MI 48109, USA}
\email{liac@umich.edu}

\author[0000-0001-8470-749X]{Elisa Costantini}
\affiliation{SRON Netherlands Institute for Space Research, Leiden, The Netherlands}
\email{e.costantini@sron.nl}

% ...



\begin{abstract}
The XRISM/Resolve microcalorimeter directly measured the gas velocities in the core of the Virgo Cluster, the closest example of AGN feedback in a cluster. This proximity allows us to resolve the kinematic impact of feedback on scales down to 5\,kpc. Our spectral analysis reveals a high velocity dispersion of $\sigma_v=262^{+45}_{-38}$ km/s near the AGN, which steeply declines to $\sim$60\,km/s between 5 and 25\,kpc in the northwest direction. The observed line-of-sight bulk velocity in all regions is broadly consistent with the central galaxy, M87, with a mild trend toward blueshifted motions at larger radii. Systematic uncertainties have been carefully assessed and do not affect the measurements. The central velocities, if attributed entirely to isotropic turbulence, correspond to a transonic ICM at sub-6 kpc scales with three-dimensional Mach number $0.69^{+0.14}_{-0.11}$ and a non-thermal pressure fraction of $21^{+7}_{-5}\%$. Simple models of weak shocks and sound waves and calculations assuming isotropic turbulence both support the hypothesis that the velocity field reflects a mix of shock-driven expansion and turbulence. Compared to other clusters observed by XRISM to date, M87's central region stands out as the most kinematically disturbed, exhibiting both the highest velocity dispersion and the largest 3D Mach number, concentrated at the smallest physical scales.
\end{abstract}


\section{Introduction}
The Virgo Cluster, with galaxy M87 at its center, is the closest example of AGN feedback in a cluster of galaxies. This trait allows for the resolution of smaller scales than is possible in any other massive halo, and has made Virgo/M87 a prime object for the X-ray study of feedback and galaxy clusters. It has been observed with all modern X-ray missions, including ROSAT, XMM-Newton, Chandra, Suzaku, and eROSITA \citep[e.g.,][]{Boehringer1994, Boehringer2001, Belsole2001, Churazov2001, Young2002, Forman2005, Forman2007, Simionescu2007, Urban2011, Arevalo2016, Simionescu2017, McCall2024}. These studies uncovered a variety of processes within the Virgo Cluster, including an AGN with an active jet \citep{Sparks1996, Harris2003} responsible for the inflation of a series of radio-filled bubbles that appear as cavities in X-ray images; the so-called ``arms’’, bright in soft X-rays, that are thought to be cooler gas uplifted by buoyantly rising bubbles; and a collection of high pressure regions around the AGN likely due to shocks, including a prominent, nearly circular shock front located at 2.8'. Beyond the innermost regions, the Virgo Cluster hosts a series of cold fronts in a spiral pattern characteristic of sloshing induced by mergers \citep{Markevitch2007, Werner2016}. 

Prior works have directly examined the distribution of temperature and metal abundance in the cluster, but due to the limitations of CCD detectors, it has only been possible to acquire upper limits or use indirect methods to probe the large-scale kinematic structure of Virgo. Notably, \cite{Zhuravleva2014} used a method applied to Chandra observations that uses a relationship between the amplitude of density fluctuations and the velocity of gas motions to estimate the one-dimensional velocity in M87, finding that turbulent heating is sufficient to offset radiative cooling. \cite{Pinto2015} reported upper limits on velocity broadening in M87 using XMM-Newton RGS spectroscopy. Additionally, \cite{Gatuzz2022} used a technique presented by \cite{Sanders2020} that uses background X-ray lines in XMM-Newton EPIC-pn to map the bulk velocity distribution in Virgo/M87, finding evidence that both AGN outflows and gas sloshing contribute to the velocity structure.

Only with the advent of X-ray microcalorimeters, such as the recently launched XRISM/Resolve, can we directly measure gas velocities in galaxy clusters. The high-resolution X-ray spectroscopy achievable with Resolve allows us to probe bulk velocity and velocity dispersion via the shift and broadening of prominent emission lines. In this work, we present a kinematic analysis of two XRISM/Resolve observations focused on the central and northwest regions of Virgo/M87 with the aim of directly measuring velocities in the area of AGN feedback. These observations are part of a larger campaign that also observed the arms of M87, which will be the focus of a future work.

In Section 2, we introduce the XRISM/Resolve observations and the supporting Chandra and NuSTAR observations that aided in constraining the non-thermal components of the spectral fit. Section 3 contains an explanation of the data analysis steps required to arrive at a final best-fit spectrum, from which we extracted the velocity measurements presented in Section 4. We discuss the interpretation of the findings in Section 5 and finish with a summary of the work in Section 6.

Throughout this paper, we use the cosmological parameters $H_0 = 70$\,km s$^{-1}$ Mpc$^{-1}$, $\Omega_m = 0.3$, and $\Omega_{\Lambda} = 0.7$. Thus, $1'$ corresponds to $4.65$\,kpc at the 
distance of 16.7\,Mpc \citep{Mei2007}. 
Quoted uncertainties are 1$\sigma$ unless otherwise stated.

\section{Observations and Data Reduction}

\subsection{XRISM/Resolve}
\subsubsection{Observations} 

M87 was observed six times with XRISM/Resolve: five times between May 26th and June 15th, 2024, and a sixth time between December 9th and 11th, 2024. All observations were conducted with the open filter wheel configuration and with the gate valve closed, limiting the soft response. This work focuses on a subset of the data collected in May and June 2024: one observation (obsID 300014010) pointed at the cluster center (RA=187.704,DEC=12.389) for 116\,ks and two observations (obsIDs 300016010 and 300016020) pointed to the northwest (RA=187.678,DEC=12.435) for 26\,ks and 143\,ks, respectively, for a total of 169\,ks. This allows us to probe the radial gradients of the bulk velocity and velocity dispersion towards a relatively relaxed azimuth, away from the E and SW ``arms'', which we defer to a future publication.  

\subsubsection{Data Reduction} \label{subsec:xrism_reduc}

The XRISM observation data were processed using pre-release Build 8 XRISM pipeline software and calibrated with the version 8 CALDB, with the exception of the CalDB redistribution matrix parameter file \texttt{xa\_rsl\_rmfparam\_20190101v006.fits}. This file is included in subsequent versions of the CalDB and includes updated FWHM parameters of the cores of the high-resolution primary (Hp) line-spread functions based on in-flight calibration. The standard event screening criteria were applied following the XRISM ABC Guide version 1.0\footnote{\url{https://heasarc.gsfc.nasa.gov/docs/xrism/analysis/abc_guide/xrism_abc.html}}, including post-pipeline screening on rise time and pixel-to-pixel coincidence to minimize the particle-induced background. Throughout the analysis, pixel 27 was excluded due to its frequent energy-scale jumps that cannot be corrected by the energy-scale tracking procedure currently in use \citep{Porter2024}. Pixel 11 also experiences abrupt changes, but infrequently.  We identified such a jump in the energy scale of pixel 11 in obsID 300016020, via comparison to the gain trend on the other pixels provided in the energy-scale calibration report\footnote{\url{https://heasarc.gsfc.nasa.gov/docs/xrism/analysis/gainreports/index.html}} for that obsID, and have therefore also excluded pixel 11 in that data set.  To account for false low-resolution secondary (Ls) events associated with clipped pulses, an event file with all Ls events removed and the screening \texttt{PI>=4000} and \texttt{PI<=20000} was used to create the RMF files. 

Spectra were extracted from portions of the total field-of-view (FOV) using only Hp events; the region selection is explained in subsection~\ref{subsec:SSM}. All spectra were then optimally binned \citep{Kaastra2016} using the \texttt{ftgrouppha} tool. For each region, we created exposure maps, redistribution matrix files (RMFs), and auxiliary response files (ARFs) using the \texttt{xaexpmap}, \texttt{rslmkrmf}, and \texttt{xaarfgen} tools. Size large RMFs were used, which are sufficient for spectral fitting that excludes energies below 3 keV. 

To construct ARFs that capture the spatial distribution of the ICM, we used exposure-corrected 3.0-7.0\,keV Chandra ACIS images of M87 as input to the \texttt{xaarfgen} tool in IMAGE mode.
This band was chosen because of its significant overlap with the fitted spectral range. 
Since the goal was to capture only ICM photons in ARF generation, the M87 AGN and jet were masked from the Chandra image and filled in with pixel values sampled from the surrounding region. The ARFs for the AGN and jet, and for the low-mass X-ray binaries (LMXBs) were generated in POINT SOURCE mode. Details of the Chandra data reduction are in section~\ref{sec:chandra}, while treatment of non-thermal emission and how the ARFs were used in spectral modeling can be found in section~\ref{sec:results}.

We reviewed the energy scale calibration reports for the three obsIDs. Correcting the energy scale of the dedicated calibration pixel, which is continuously illuminated outside of the aperture, based on the same calibration time intervals as the main array, makes it a witness for the goodness of the intermittent calibration for the main array.  We found the worst line offset, 0.19 eV, in OBSID 300014010.  Adding in quadrature with the current energy-scale accuracy from 5.4--9.0-keV of $\pm0.3$~eV results in $\pm0.35$~eV, and $\sim$0.3 eV for all three pointings, resulting in a systematic line-of-sight bulk-velocity uncertainty of $<18$ kms$^{-1}$ at 6 keV.

Uncertainty in the instrumental line widths comes from a combination of measurement uncertainty, time-dependent noise, and broadening induced by the time-dependent energy-scale correction.  The calibration-pixel FWHM widths ranged from 4.40 to 4.50 eV in these observations.  The current in-orbit calibration has a line-spread uncertainty of $\sim0.15$ eV from 6--7 keV.  This error at 6 keV results in a systematic uncertainty in the turbulent velocity of  $\pm5\,{\rm km\,s^{-1}}$ for turbulence of 60 km\,s$^{-1}$ and $\pm1.2\,{\rm km\,s^{-1}}$ for turbulence of 260 km\,s$^{-1}$.  Additional line broadening can come from energy-scale misalignment of different pixels or observations. Preliminary results of a recent calibration investigation indicate that the error at 6.5 keV across the pixels is $\sim$0.2 eV and for the whole array across different integration intervals is $\sim$0.02 eV. The larger of these represents a broadening of 9 km\,s$^{-1}$ at 6.5 keV, but since it is a noise term that adds in quadrature with the actual velocity dispersion, it represents an additional uncertainty of $<$1 km\,s$^{-1}$ atop an inferred turbulence of 60 km\,s$^{-1}$.

The non-X-ray background (NXB) spectra were estimated from the first version of Resolve's NXB database, built from observations of the dark Earth, using the \texttt{rslnxbgen} task. To generate NXB spectra for the desired pixels in each observation, background events were selected based on the cut-off rigidity distribution of the observation, and the same event screening criteria as for the source spectrum were applied. The typical NXB level in the regions considered in this analysis is of the order $10^{-4}\, \textrm{counts}\ \textrm{s}^{-1}\ \textrm{keV}^{-1}$.

\subsection{Chandra}\label{sec:chandra}

\subsubsection{Observations}

M87 has been regularly observed by \textit{Chandra} to monitor the AGN and jet variability \citep{Harris2006}, and the extended hot atmosphere was the target of deep observations in 2005 \citep{Forman2007}.  AGN monitoring observations use a subarray and short frame-time to minimize pileup due to the bright nucleus.  This allows us to accurately extract the spectral properties of the AGN but restricts the field of view to an arcmin-wide strip across the nucleus.  The hot atmosphere's structure is revealed by the full field of view but the AGN emission is so heavily piled up that affected events were not telemetered to the ground.  This produces clear detector artifacts, including a hole at the center of the AGN's PSF and a strong readout streak.  Here we utilise the monitoring observations taken on 8 April 2024, close to the XRISM observations, to determine the AGN spectrum (Obs. IDs 26761 and 26762, 4.7 and 4.6\,ks long respectively), longer monitoring observations taken in 2016 (Obs. IDs 18232, 18233, 18781, 18782, 18783, 18836, 18837, 18838, 18856, totaling $298\,$ks) and 2022 (Obs. ID 25369, 34.3\,ks) for comparison, and deep full field observations to map the hot atmosphere and measure the LMXB contribution, excluding the AGN and detector artefacts (Obs. IDs 352, 2707, 3717, 5826, 5827, 5828, 6186, 7210, 7211 and 7212).

\subsubsection{Data Reduction}

All \textit{Chandra} observations were reprocessed and analysed with \textsc{ciao} version 4.17 and \textsc{caldb} version 4.11.6 provided by the \textit{Chandra X-ray Center} \citep{Fruscione2006}.  For details of the reprocessing, flare cleaning and blank sky background generation, see \citet{Russell2018}.  Exposure-corrected images for the full field in the energy band $3-7$\,keV were produced from each individual observation and summed together.

\subsection{NuSTAR}

\subsubsection{Observations}

To better understand the behavior of the AGN, we also used a Nuclear Spectroscopic Telescope Array (NuSTAR) \citep{Harrison2013} observation of M87 from April 9th of 2024 (Obs. ID 60802031004), which is close in observation time to the XRISM observations. During the $\sim$55 ks observation, both focal plane modules (i.e. FPMA and FPMB) were used.

\subsubsection{Data Reduction}

In order to filter the data, standard pipeline processing using HEASoft (v.~6.34) and NuSTARDAS (v.~2.1.4a) tools were used with the CALDB released on 22 January 2025. Stage 1 and 2 of the NuSTARDAS pipeline processing script \textsc{nupipeline} were utilized to clean the data. Regarding the cleaning of the event files for the passages through the South Atlantic Anomaly (SAA) and a ``tentacle"-like region of higher activity near part of the SAA, instead of using SAAMODE=STRICT and TENTACLE=yes calls, we produced light curves and manually removed increased count rate intervals using 100 s time bins to create good time intervals (GTIs) without fully discarding the passage intervals.

This new set of GTIs was then reprocessed with \textsc{nupipeline} stages 1 and 2, and cleaned images were generated at different energy bands with XSELECT. We used \textsc{nuexpomap} to create exposure maps in the 3.0~--~8.0 keV and 8.0~--~15.0 keV bands accounting for vignetting. To produce the corresponding spectra for the regions of interest, as well as the corresponding Response Matrix Files (RMFs) and Ancillary Response Files (ARFs), stage 3 of the \textsc{nuproducts} pipeline was used.

We modeled the NuSTAR background using a set of IDL routines called \textsc{nuskybgd} that have become a standard for background assessment of extended sources in the last decade. The treatment of the background and its components are explained in detail in \citet{Wik2014}. We applied \textsc{nuskybgd} to the spectra extracted from regions that are mostly free of cluster emission, but still account for residual ICM emission by using a single temperature model.

\begin{figure*}
 \centering
  \includegraphics[width=0.8\linewidth]{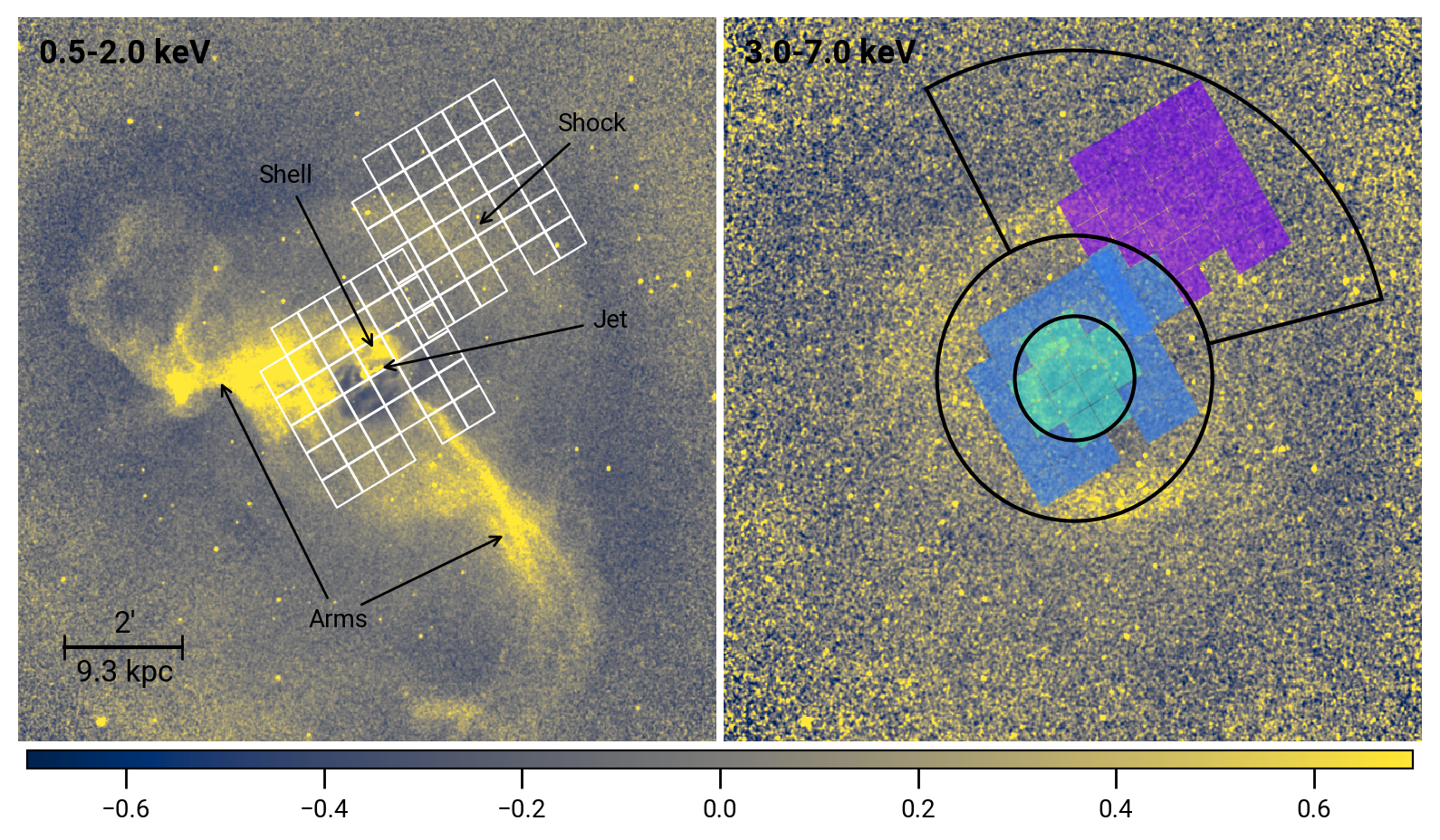} 
\caption{Residual Chandra images (images divided by the best-fit models to emphasize substructure) of M87 with the central and northwest Resolve pointings overlaid. \textit{Left:} 0.5-2.0 keV image. In this energy band, M87's AGN/jet and arms are prominent. \textit{Right:} 3.0-7.0 keV image. The AGN and jet have been masked and filled in with values of surrounding pixels. The division into 3 sky regions is illustrated by the black shapes, with the corresponding detector regions shown in color. This energy band was used for image ARF creation.}\label{fig:image}
\end{figure*}

\section{Data Analysis} \label{sec:results}

\subsection{Non-thermal contribution from the AGN and LMXBs} \label{subsec:nontherm}
Due to the mixing of the continuum ICM emission and the power law-like AGN/jet and LMXB contributions, XRISM data alone was not sufficient to constrain all parameters simultaneously. Chandra and NuSTAR data were used to better understand the non-thermal contribution from the AGN/jet and LMXBs.

\subsubsection{Chandra}

A total spectrum for the AGN and jet was extracted from each of the two monitoring observations using a circular region with 2\arcsec\ radius centered on the AGN and a rectangular region covering the jet with dimensions $18\arcsec \times3\arcsec$.  The cluster background was extracted from an annulus centered on the AGN from 3 to $5\arcsec$ radius and restricted to $65-330\,^\circ$ to avoid the jet emission.  Appropriate responses were generated and the spectra were grouped to a minimum of 1 count per bin.  The spectra were fit together in \textsc{xspec} version 12.14.1 \citep{Arnaud1996} over the energy range $0.5-7\,$keV with an absorbed power law model \textsc{TBabs(zTBabs(powerlaw))}.  The Galactic absorption was fixed to $1.26\times10^{20}\,\textrm{cm}^{-2}$ \citep{HI4PI}, the redshift was fixed to $z=0.00428$ and the intrinsic absorption column density, photon index and normalization were left free.  The best-fit values were determined by minimizing \textsc{xspec}'s version of the C-statistic \citep{Cash1979,Wachter1979}.  

The best-fit intrinsic absorption is $n_{\mathrm{H,z}}=0.2\pm0.1\times10^{22}\,\textrm{cm}^{-2}$, the photon index $\Gamma=2.4\pm0.1$ and the unabsorbed flux in the energy range $2-10\,$keV is $1.52^{+0.10}_{-0.09}\times10^{-12}\,\textrm{erg}\,\textrm{cm}^{-2}\,\textrm{s}^{-1}$ (and $1.27^{+0.07}_{-0.06}\times10^{-12}\,\textrm{erg}\,\textrm{cm}^{-2}\,\textrm{s}^{-1}$ for $2-7\,$keV).  The flux shows a modest decrease since the longer monitoring observation taken in 2022 with $1.80\pm0.04\times10^{-12}\,\mathrm{erg}\,\mathrm{cm}^{-2}\mathrm{s}^{-1}$.  The photon index is consistent with the values measured in 2016 and 2022, for which $\Gamma=2.35\pm0.02$ \citep{Russell2018}.  The intrinsic absorption is also not significantly above earlier upper limits of $n_{\mathrm{H,z}}\leq 0.02\times10^{22}\,\textrm{cm}^{-2}$ ($3\sigma$), when correcting for the difference in Galactic absorption values between these studies.  The variation in nuclear flux is consistent with the pattern of variability seen over the \textit{Chandra} mission \citep{Harris2009,Russell2018}.  For the nuclear component in the XRISM analysis, we therefore fix the spectral parameters to $\Gamma=2.35$ and 2-7 keV unabsorbed flux = $1.3 \times 10^{-12}\ \textrm{erg}/\textrm{cm}^2/\textrm{s}$. We explore the effects of these parameter choices further in subsection~\ref{subsec:systematics}.

We estimate the LMXB contribution by extracting the combined spectrum of resolved point sources in a deep full field observation (obs. ID 5827 taken in 2005) and tracking variability with the partial view provided by monitoring observations. We considered a $1\arcmin$ radius region of a broad band image where the AGN, jet, and readout streak were excluded. LMXBs were then identified with \textsc{wavdetect} \citep{Freeman2002} and confirmed by eye. A circle of radius 2\arcsec was used for each LMXB.   The cluster background emission is particularly bright and was also evaluated within a region of radius 1\arcmin\ centered on the AGN (excluding LMXB regions).  We extracted a combined spectrum for the LMXBs and a spectrum for the cluster emission and generated appropriate responses.  The LMXB spectrum was fitted with a powerlaw model, with fixed photon index $\Gamma=1.65$ \citep{Revnivtsev2014}, and a model for the cluster background.  The cluster background model was determined by fitting a \textsc{TBabs(vapec+vapec)} model to the cluster spectrum.  The best-fit parameters from this background model were fixed, with the exception of the normalizations.  The 2-10 keV LMXB flux in 2005 was $\sim3\times10^{-13}\ \textrm{erg}/\textrm{cm}^2/\textrm{s}$ (and 2-7 keV flux $\sim2\times10^{-13}\ \textrm{erg}/\textrm{cm}^2/\textrm{s}$).  Based on a partial view of the LMXB population (region with radius $0.6\arcmin$) in more recent monitoring observations from 2022 to 2024, we estimate that the LMXB flux may have dropped by as much as a factor of 3 since 2005.  The dominance of the cluster emission prevents an evaluation of the unresolved LMXB contribution.

\subsubsection{NuSTAR}

NuSTAR spectra and response files for both FPMs were extracted for a 1$\arcmin$ radius region centered at the image emission peak using the \textsc{nuproducts} script; the corresponding background was extracted using the \textsc{nuskybgd} routine. The NuSTAR spectra were grouped to have at least 3 counts per bin.

The extracted spectra were fit in the $3.0$-$11.0\,$keV and $3.0$-$15.0\,$keV energy bands using \textsc{xspec} version 12.14.1 and AtomDB 3.0.9. The former energy band is used to provide a direct comparison for the Resolve fits and the latter energy band was used to take full advantage of NuSTAR's large effective area at hard energies. Using C-statistics, we fit the data with a model \textsc{TBabs$\times$((cflux$\times$apec))+(cflux$\times$powerlaw))} to account for both the thermal emission from the ICM and AGN emission. This model is convolved with a \textsc{constant} model to account for the cross-calibration between FPMA and FPMB. 
The \textsc{constant} model accounts for the cross-calibration between FPMA and FPMB. We find that both instrument fluxes agree within 1\%. Galactic absorption is too low to have an effect on the observables in the energy bands we use, but it was modeled for consistency with other instruments. For both energy bands, we performed the fits with fixed redshift ($z=0.00428$) and free power-law photon index, and then with fixed redshift and fixed power-law photon index. The fixed power-law photon index was the Chandra-obtained $\Gamma = 2.35$ to compare what the flux would be if the instruments detected the same power-law slope. Here we note that the powerlaw is truly representing the emission 
from both the AGN and LMXB, since NuSTAR data were not able to constrain two separate powerlaw model parameters. 

We calculated the ICM flux in the $4.0$-$6.0\,$keV band and the AGN flux in the $3.0$-$7.0\,$keV band for each of our $3.0$-$11.0\,$keV and $3.0$-$15.0\,$keV fits. 
All of the steps of the fit are repeated by accounting for the gain shift of NuSTAR using the \textsc{xspec} \textsc{gain} command. We found that the gain offset was around -0.1 keV as reported in literature \citep{Rojas23}. We present the results in Table \ref{tab:nustarfit}.

\begin{table*}
  \begin{tabular}{cccccc}
      \hline
      Energy range (keV) & LMXB? & Pointing & Temperature (keV) & $\sigma$ (km/s) & Redshift $z$ (10$^{-3}$) \\
      \hline
      3-11 & No & C & 2.10$^{+0.03}_{-0.02}$ & 171$\pm 11$ & 4.36$\pm 0.03$\\
      3-11 & No & NW & 2.30$^{+0.03}_{-0.04}$ & 62$^{+14}_{-16}$ & 4.25$\pm 0.02$\\
      3-11 & Yes & C & 1.87$^{+0.04}_{-0.03}$ & 164$\pm 11$ & 4.34$\pm0.03$\\
      3-11 & Yes & NW & 2.13$^{+0.08}_{-0.05}$ & 54$^{+15}_{-18}$ & 4.24$\pm 0.02$\\
      4.2-7\footnote{Although 4.2-7 keV was chosen as the default fit, any narrow band fit including the Fe lines is dominated by their contribution and yields results in agreement with that fit.} & No & C & 1.90$^{+0.06}_{-0.04}$ & 158$^{+16}_{-15}$ & 4.42$\pm 0.04$\\
      4.2-7 & No & NW & 2.32$^{+0.07}_{-0.06}$ & 67$^{+15}_{-17}$ & 4.28$\pm 0.03$\\
      4.2-7 & Yes & C & 1.80$\pm 0.04$ & 153$\pm 16$ & 4.40$\pm0.04$\\
      4.2-7 & Yes & NW & 2.24$^{+0.07}_{-0.06}$ & 62$^{+16}_{-18}$ & 4.27$\pm0.03$ \\
      \hline
    \end{tabular}
\caption{Best-fit values for the full central (C) and northwest (NW) pointings without the inclusion of spatial-spectral mixing or flux ratio-linking. For the 4.2-7 keV fit including LMXB, the LMXB fluxes are frozen to their 3-11 keV fit values. All fits include NXB, and all C fits include the AGN power-law component with its parameters frozen.}\label{tab:fullresults}
\end{table*}

\subsection{XRISM data analysis}
We began our analysis with a straightforward approach, using the full central and northwest pointings as our regions. The results for this simplified analysis can be found in Table~\ref{tab:fullresults}. In the following sections, we describe in more detail our choice of smaller regions and the steps required to model them properly.

\subsubsection{Spatial-Spectral Mixing} \label{subsec:SSM}

To maximize our knowledge of the ICM velocity structure, we split the two pointings into three annular ``sky'' regions. The choice of regions was driven by (i) the XRISM/Resolve PSF ($\sim$1.3\arcmin), which sets the approximate minimum annulus width and (ii) the need for enough photon counts in each region such that velocity broadening can be measured. These requirements motivated the choice of (i) region 1, a circle of 1\arcmin\ radius containing the central 12 pixels of the central pointing, (ii) region 2, an annulus of outer radius 2.3\arcmin\ that contains the rest of the central pointing and 5 pixels of the northwest pointing and (iii) region 3, an annulus of radius 5.3\arcmin\ that encompasses the remainder of the northwest pointing. The pointings and region divisions can be seen overlaid on Chandra ACIS residual images in Figure~\ref{fig:image}.

Since the Half Power Diameter (HPD) of XRISM/Resolve is 1.3\arcmin, the X-ray spectrum of a given region contains contamination of photons from neighboring regions that must be accounted for in the spectral modeling. This is done via the so-called Spatial-Spectral Mixing (SSM) analysis, in which the fraction of photon leakage between regions is evaluated using ray-tracing to produce the ARFs. Using the \texttt{xrtraytrace} task, we simulated the paths of photons from neighboring sky regions (with the initial photon distribution assumed to follow the Chandra image described in subsection~\ref{subsec:xrism_reduc}) onto the detector, accounting for the shape of Resolve's PSF. The model fit to the detector region spectrum then assumes contributions from all sky regions. Fits were initially performed using all available XRISM/Resolve data of M87 to check the photon contribution from all regions within a 5.3\arcmin\ radius of the central AGN. As their effect on the fit parameters was negligible, however, regions contributing $< 5\%$ of the total photons in a detector region were not included in the default fit; the eastern and southwestern region contributions were therefore not required. SSM of the central AGN and jet, and the LMXB component from each region were modeled the same way as the ICM, but using point source ARFs rather than image ARFs. 

\subsubsection{Spectral modeling} \label{subsec:model}

We fit Resolve spectra using \textsc{xspec} 12.15.0 (\cite{Arnaud1996}), with corresponding atomic database AtomDB 3.1.3. All abundances are relative to \citet{Lodders2009} proto-Solar abundances. The best-fit values were determined by minimizing \textsc{xspec}'s version of the C-statistic \citep{Cash1979,Wachter1979}.  

The NXB, extracted for each detector region in each observation as described in section~\ref{subsec:xrism_reduc}, was modeled separately from and prior to the M87 dataset, in the 3-11\,keV band.
Its model consists of a single power-law and Gaussian instrumental emission lines. 
The normalization of the overall NXB for each spectrum was fit first, followed by the normalizations of the individual emission lines. The best-fit model for the NXB contribution was then frozen for all subsequent fits.

The ICM emission from each region was modeled as a collisionally-ionized, single temperature thermal plasma with thermal and velocity broadening (\texttt{bvapec} model in XSPEC). 
For each region, column density $n_{\mathrm{H}}$ was fixed to the weighted average $1.26 \times 10^{20}\ \textrm{cm}^{-2}$ \citep{HI4PI} and abundances of He and C were fixed to Solar. 

In addition to the three separate ICM components, in region 1 the AGN and its jet were modeled with an absorbed power-law with 2-7 keV flux $1.3 \times 10^{-12}\ \textrm{erg}/\textrm{cm}^2/\textrm{s}$ and photon index 2.35, which were both constrained by Chandra data where the AGN and jet are spatially resolved. The intrinsic absorption was neglected because it was found not to have a strong effect above 3 keV, and thus its inclusion did not change any best-fit parameters. Since the spectral shape of the AGN is known not to be strongly varying \citep{Russell2015, Russell2018}, these parameters were frozen. Each of the three regions also included a power-law model for the contribution of the LMXBs with photon index frozen to 1.65 (e.g., \cite{Irwin2003, Revnivtsev2014}). Estimates of the resolved LMXB fluxes in each region were taken from existing Chandra data. However, as LMXB fluxes are variable with time and Chandra observations are not simultaneous, the LMXB fluxes were left as free parameters and determined by the XRISM data. The Chandra measurements were used only to assess whether the findings were reasonable.

\begin{figure}
    \centering
    \includegraphics[width=\linewidth]{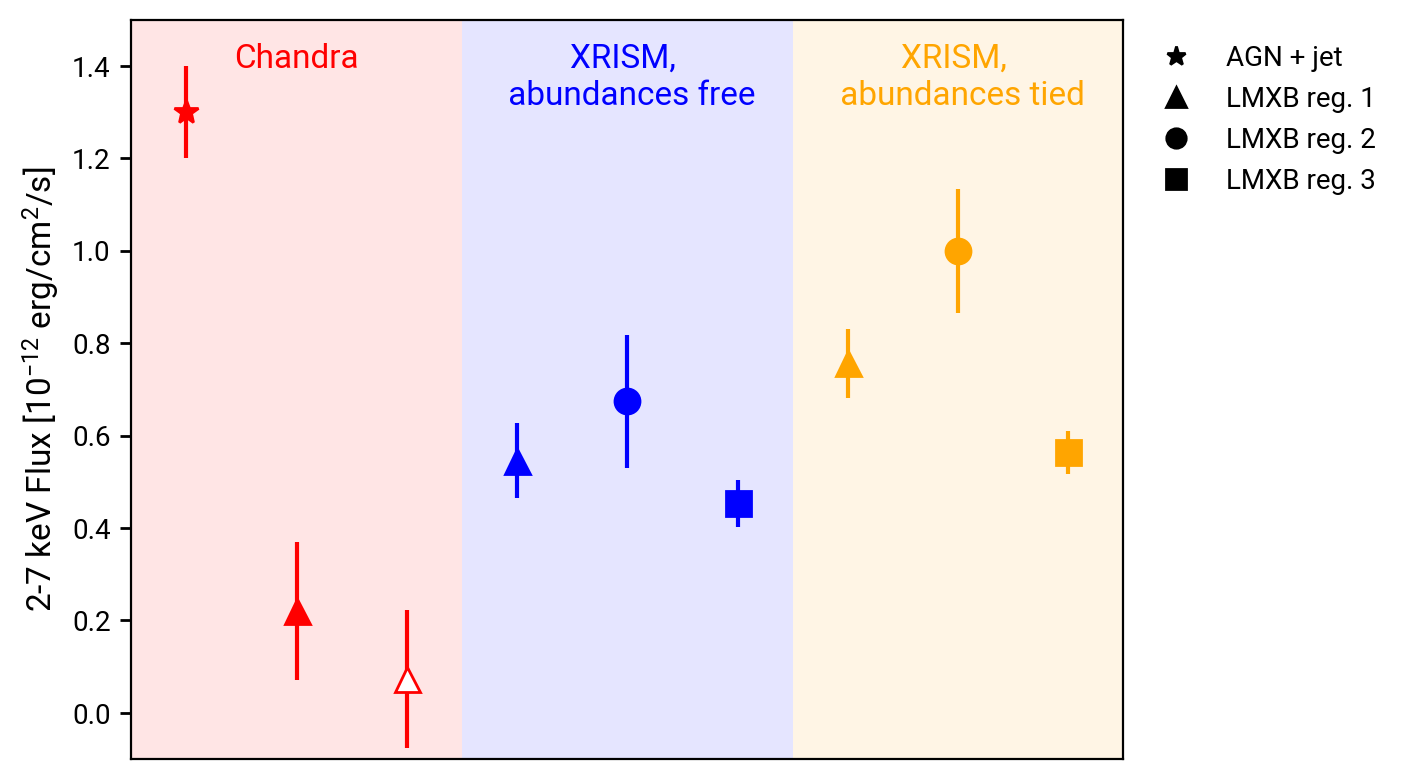}
    \caption{Flux comparison between best-fit values from Chandra (red), XRISM with abundances free (blue), and XRISM with abundances tied to Fe (orange). The red-filled triangle point comes from resolved LMXBs in 2005 Chandra data. The white-filled triangle point comes from a value taken from 2024 Chandra data within a 0.6’ radius, extrapolated to the expected value for a 1’ radius for comparison with detector 1. XRISM fits were performed in the band 3-11 keV.}
    \label{fig:lmxb}
\end{figure}

When both ICM and LMXB fluxes are left free for each region, LMXB fluxes become unreasonably high, preferring, for example, a 2-7 keV flux in region 1 that is larger than the contribution from the AGN and jet. From Chandra estimates, the LMXB contribution in this region should be 20-30\% that of the AGN and jet (see Figure~\ref{fig:lmxb}). In order to provide an additional constraint,  we find the continuum fluxes (4-6 keV) from the 4-7 keV fit to the Chandra ICM in each sky region (first removing LMXBs and AGN/jet), and take the ratio of these fluxes (i.e., sky1/ sky2 and sky2/sky3). In the XRISM fit, we fix the ratio between the ICM regions to the values from Chandra, assuming that the relative ICM fluxes (not their absolute values) should remain the same regardless of instrument.

To then find the best-fit LMXB fluxes, we performed a fit in the broad band, 3-11 keV. In this model, the gas temperature, metallicity (of elements S, Ar, Ca, Fe, and Ni), redshift, and Gaussian sigma of the velocity broadening were treated as free parameters, in addition to the 2-7 keV LMXB flux of each region. The 4-6 keV fluxes of the ICM components were tied via their Chandra ratio values to provide more constraints to the fit, as described above. This approach resulted in LMXB fluxes of the same order of magnitude as the Chandra estimates, as can be seen in Figure~\ref{fig:lmxb}. This figure provides a visual comparison of the flux of the AGN from Chandra, of the resolved LMXBs in region 1 from Chandra, and of the LMXBs in each region from XRISM fits. The LMXB fluxes from two different XRISM fits are shown; one is the 3-11 keV fit described in this section, and the other is a 3-11 keV fit where all abundances are tied to Fe, the motivation for which is discussed further in subsection~\ref{subsec:systematics}.

After the LMXB fluxes were obtained in the 3-11 keV fit, their values were frozen for the fit henceforth referred to as the ``default’’, in the 4.2-7 keV band. This band was chosen because it is dominated by Fe emission lines and excludes prominent lines from other elements. For this fit, the gas temperature, Fe metallicity, redshift, and Gaussian sigma of the velocity broadening were treated as free parameters. All other element abundances were tied to Fe. The ICM components were again tied to their Chandra ratio.

\begin{figure*}
 \begin{center}
  \includegraphics[width=0.99\linewidth]{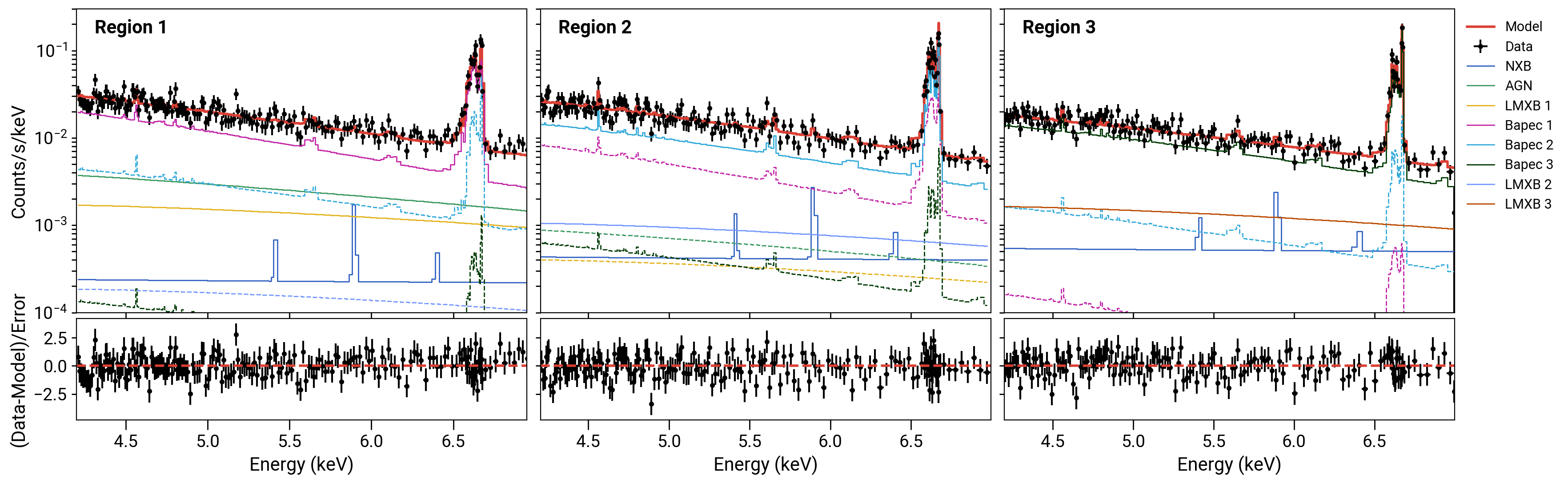} 
  \includegraphics[width=\linewidth]{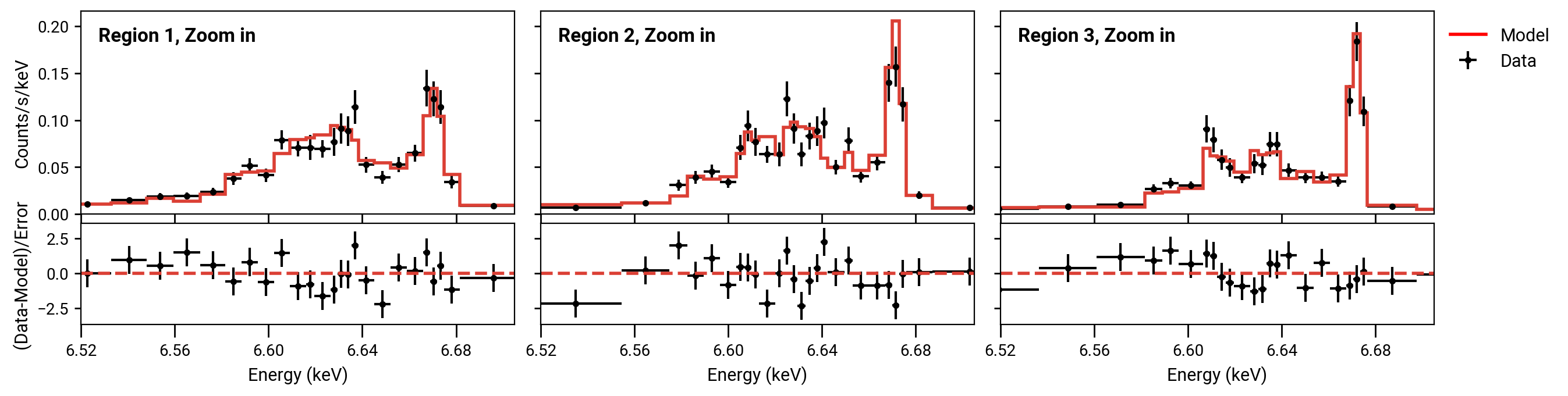}
 \end{center}
\caption{Observed spectra, best-fit models, and fit residuals using default choice of modeling (see subsection~\ref{subsec:model}). \textit{Top panels:} 4.2-7 keV fit for each of the 3 regions, including components that contributed to the total model. Contributions from off-set regions due to spatial spectral mixing are plotted as dashed lines. \textit{Bottom panels:} Zoomed view of the strongest He-like Fe emission lines and fit residuals from the 4.2-7 keV fit.}\label{fig:spec}
\end{figure*}

\section{Results}

\begin{figure*}
    \centering
    \includegraphics[width=0.8\linewidth]{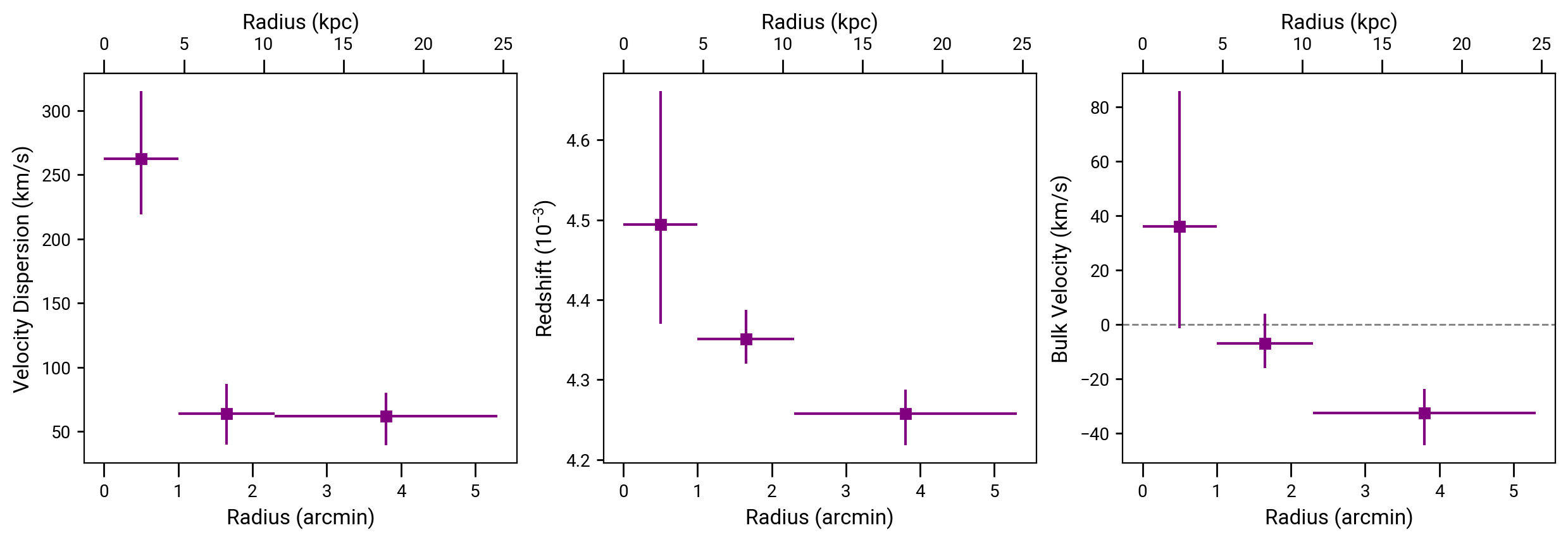}
    \caption{Distribution of the velocity-related properties of the ICM from the default 4.2-7 keV fit, as described in the text. Note that the redshifts are not heliocentric corrected, as this step is performed in the conversion to bulk velocity.}
    \label{fig:defaults}
\end{figure*}

Figure~\ref{fig:spec} shows the data and best-fit models for the 4.2-7 keV spectra, including the contributions from relevant components, and zoom-ins to the He-like Fe lines. 
Contributions from external regions, accounted for with SSM modeling, are shown as dashed lines. Figure~\ref{fig:defaults} shows the radial profiles of the velocity-related measurements for the default fit.
There is a large velocity dispersion of $\sim 265$ km/s in region 1 (within a radius of 5 kpc around the AGN) and a steep decrease from region 1 to 2, down to $\sim 60$ km/s. 
The trend in the redshift is much weaker, with the gas becoming less redshifted from region 1 to region 3.
Table~\ref{tab:results} summarizes the best-fit parameters from the default fits and their statistical and systematic errors. The systematic uncertainties are explained in more detail in subsection~\ref{subsec:systematics}.

We determine the bulk velocity from the line shifts, given by
\begin{equation}
v_{\mathrm{bulk}} = \frac{c(z - z_{BCG})}{1 + z_{BCG}}
\end{equation}
where $c$ is the speed of light, $z$ is the cosmological redshift, and $z_{BCG}$ is the redshift of the BCG, 0.00428 \citep{Cappellari2011}.
The barycen heasoft tool, which uses the orbit auxiliary files, was used to obtain the heliocentric corrections. 
This correction varies from +27.99 to +28.16 km/s across the duration of the central pointing, and from +25.25 to +26.10 km/s across the two northwest pointings. For simplicity, an average correction of +28.1 km/s and +25.6 km/s was applied to the C and NW regions, respectively. 

\subsection{Systematic Uncertainties} \label{subsec:systematics}

\begin{figure*}
    \centering
    \includegraphics[width=\linewidth]{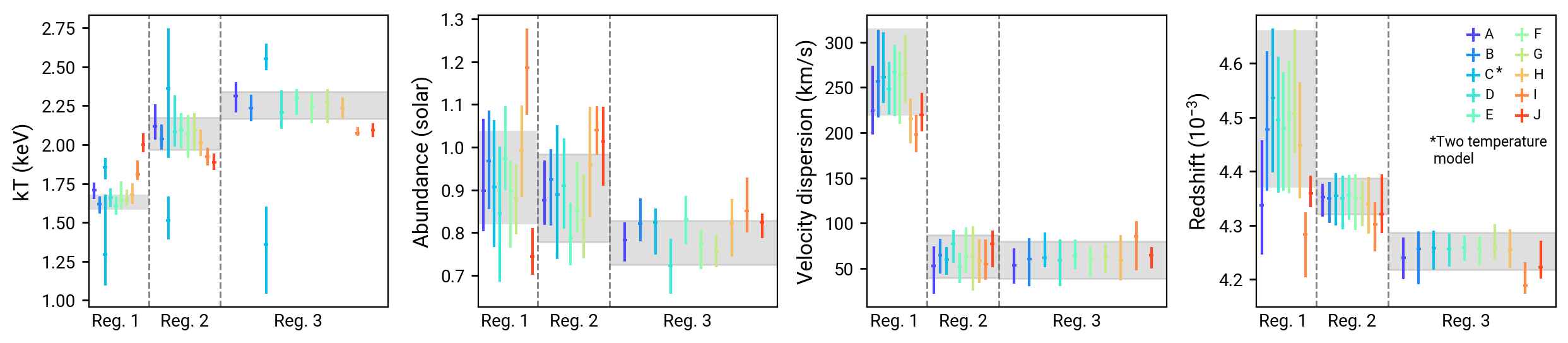}
    \caption{Distribution of the measured properties of the ICM shown with 1$\sigma$ uncertainties. The four panels show (1) gas temperature, (2) Fe abundance with respect to solar, (3) line-of-sight velocity dispersion, (4) redshift. The gray shaded regions demarcate the statistical uncertainties from the default 4.2-7 keV band fit. The points, from left to right, correspond to (A) the 4.2-7 keV band but calculated with AtomDB v. 3.0.9, (B) the 4.2-7 keV band but using the LMXB fluxes from the model where abundances are tied to Fe, (C) the 4.2-7 keV band with a 2T model, (D, E) the 4.2-7 keV band with off-axis ARFs varied by $\pm 30\%$, (F, G) the 4.2-7 keV band with the NXB normalization varied by $\pm 20\%$, (H) the 4.2-7 keV band with the default choices calculated with SPEX 3.08.01, (I) the 3-11 keV band in AtomDB 3.1.3, with default choices but the He-like Fe W line removed, and (J) the 3-11 keV band with the default choices calculated in AtomDB 3.1.3.}
    \label{fig:systematics}
\end{figure*}

The systematic uncertainties were checked for the following factors that could affect the velocity measurements: (i) LMXB flux, (ii) multi-temperature models, (iii) SSM modeling, (iv) NXB model, (v) energy band, (vi) non-thermal photon indices, (vii) AtomDB version, (viii) atomic database, (ix) resonant scattering, (x) software version. Many of the tests described in this section are shown in Figures~\ref{fig:systematics} and \ref{fig:calculated}.

i) LMXB flux: It was found that untying the abundances from Fe in the 3-11 keV fit leads to a shoulder-like residual of $\sim 2 \sigma$ significance to the left of the He-like Fe lines at $\sim 6.57\ \mathrm{keV}$ in region 1. This was found to be due to the higher temperature of the fit in this case as compared to both the narrow band 4.2-7 keV case and the broad band 3-11 keV case where all abundances are tied to Fe, and was not resolved through the use of multi-temperature models. Despite this residual, the 3-11 keV case with abundances free was used to acquire LMXB fluxes for the narrow band fit because its LMXB fluxes better matched the Chandra estimates (see Figure~\ref{fig:lmxb}). To check the dependence of the velocity measurements on LMXB flux choice, however, the LMXB fluxes from the 3-11 keV fit with all abundances tied to Fe were also used in a narrow band fit. As can be seen from point B in Figure~\ref{fig:systematics}, this approach agrees within 1$\sigma$ with the default approach.

ii) multi-temperature models: Recent works \citep{Chatzigiannakis2025} suggest that single temperature models are a poor approximation for the ICM. We fit a double \texttt{bvapec} model with its velocities and abundances tied, but with its temperatures free to test how velocities are affected by the use of a two temperature model (point C in Figure~\ref{fig:systematics}). We also used the multi-temperature model \texttt{bvgadem}. Neither change resulted in velocities that were statistically different from the default single temperature fit.

iii) SSM modeling: To explore the impact of a systematic uncertainty in the modeling of photons that impact regions they did not originate in, we modified the offset ARFs’ effective areas by $30\%$ in either direction. This modification did not change the shape of the effective area, but rather scaled the entire curve. To test the limiting cases, we either scaled every offset ARF by $+30\%$ or every offset ARF by $-30\%$, and then performed the fit. As can be seen from points D and E in Figure~\ref{fig:systematics}, the uncertainties overlap with the defaults.

iv) NXB model: To evaluate the impact of the NXB modeling on the results, we scaled the normalization of the NXB by $\pm20\%$. This demonstrated that potential uncertainty in the NXB normalization does not impact the velocity measurements (points F and G in Figure~\ref{fig:systematics}).

v) energy band: The default energy band chosen for this paper, the 4.2-7 keV band, which is driven by the Fe emission lines, can be well described by a single temperature, single velocity, single abundance model. The broad 3-11 keV band may require more complicated modeling, and will therefore be treated in more detail in upcoming works. However, to test how robust the 4.2-7 keV velocities are to the use of different energy bands, we also include fits of the same regions in the range 3-11 keV for our systematics checks. The choice of energy band introduces the largest systematic uncertainty (points I and J in Figure~\ref{fig:systematics}).

vi) non-thermal photon indices: Although the LMXB fluxes from two different 3-11 keV fits were used to test systematic uncertainty (i), the power-law slopes of both the LMXBs and the AGN were fixed to values motivated by other works \citep{Irwin2003, Revnivtsev2014}. To test the effect of variation in the non-thermal components’ photon indices, we modified them by $\pm10\%$. Fits were run in which LMXB slopes were varied, the AGN slope was varied, and both the LMXB and AGN slopes were modified by $\pm10\%$ in the same fit. The 3-11 keV best-fit values were all in agreement with the default 3-11 keV fit (point J), and the 4.2-7 keV fit values were all in agreement with the default 4.2-7 keV fit (gray shaded regions).

vii) AtomDB version: While this work was in preparation, the AtomDB version was updated from 3.0.9 to 3.1.3. As this update affected the velocities, we include this version for comparison in Figure~\ref{fig:systematics} (point A). However, the choice of AtomDB version is not included in the computation of the overall systematic spread quoted in Table~\ref{tab:results}, as the assumption is made that the latest version is the most accurate. 

viii) atomic database: To check that our velocity measurements do not rely exclusively on our choice of atomic database, we also ran our default fit using SPEXACT 3.08.01 in XSPEC. This did not result in velocity measurement changes $>1\sigma$ as compared to the default (point H in Figure~\ref{fig:systematics}).

ix) resonant scattering:  To test whether velocity measurements are affected by resonant scattering \citep{Gilfanov1986,Churazov2010}, which mostly affects the flux and shape of the strongest W line \citep{Churazov2010}, we removed the W line from AtomDB and replaced it in the fit with a Gaussian, detaching ICM line broadening from the W-line (point I in Figure~\ref{fig:systematics}). We find no evidence of resonant scattering in the W line, which is consistent with our expectations for this exposure time.

x) software version: The default choices of Build 8 XRISM pipeline software and CALDB version 8 to produce all data products was compared with products from Heasoft version 6.34 and CALDB version 10. Best-fit values were unchanged within 1$\sigma$.

\begin{table*}
  \begin{tabular}{cccc}
      \hline
      Parameter & Reg 1 & Reg 2 & Reg 3  \\ 
      \hline
      Projected radial range, $R$ (') & 0--1 & 1--2.3 & 2.3--5.3 \\
      kT (keV) & 1.63 $\substack{+0.07 \\ -0.06} \pm 0.13$ & 2.08 $\substack{+0.12 \\ -0.09} \pm 0.08$ & 2.26 $\substack{+0.09 \\ -0.08} \pm 0.08$ \\
      Fe (Solar) & 0.91 $\substack{+0.11 \\ -0.10} \pm 0.12$ & 0.86 $\substack{+0.09 \\ -0.08} \pm 0.09$ & 0.78 $\pm0.05 \pm 0.04$ \\
      Redshift $z$ (10$^{-3}$) & 4.49 $\substack{+0.14 \\ -0.13} \pm 0.08$ & 4.35 $\pm 0.05 \pm 0.02$ & 4.26 $\pm0.04 \pm 0.02$ \\
      $v_{\mathrm{bulk}}$\ (km/s)\footnote{Calculated with respect to the BCG M87, and including heliocentric correction.} & 39 $\substack{+42 \\ -40} \pm 23$ & -7.1 $\pm 14 \pm 5$ & -33 $\pm 11 \pm 7$ \\
      $\sigma$ (km/s) & 262 $\substack{+45 \\ -38} \pm 26$ & 64 $\substack{+22 \\ -26} \pm 9$ & 62 $\substack{+19 \\ -22} \pm 8$ \\
      log10 2-7 keV LMXB flux (erg/cm$^2$/s)\footnote{Fit in the broad band and frozen for the 4.2-7 keV fit.}& -12.26 $\substack{+0.06 \\ -0.07}$ & -12.17 $\substack{+0.09 \\ -0.11}$ & -12.34 $\pm 0.05$ \\
      \hline
    \end{tabular}
\caption{Best-fit values from the 4.2-7\,keV fit. The first listed uncertainties are statistical and the second systematic. Systematic uncertainties are defined as the standard deviation of the test results discussed in subsection~\ref{subsec:systematics}.}\label{tab:results}
\end{table*}

\section{Discussion}\label{sec:discussion}

XRISM Resolve provides the first direct velocity measurement of the Virgo Cluster ICM. Regardless of systematics, the velocity dispersion peaks in the center, with a value of $\sim 260$\,km/s in the innermost $5$\,kpc. This region contains the AGN, jet, small cavities, and a high pressure region inside a shock front generated by the current AGN outburst \citep{Forman2007}. The velocity dispersions beyond $5$\,kpc out to $25$\,kpc, in the northwest direction, are consistently $\sim 60\,$km/s in both sub-regions. Region 2 contains the base of the outflowing arms, while a shock front of Mach number $M \sim 1.2$ with a radius $\sim 2.8'$ falls in region 3 \citep{Forman2007}. Note that this velocity profile does not include the pointings centered on the outflowing arms, the analysis of which will be featured in an upcoming paper.

\begin{figure*}
    \centering
    \includegraphics[width=\linewidth]{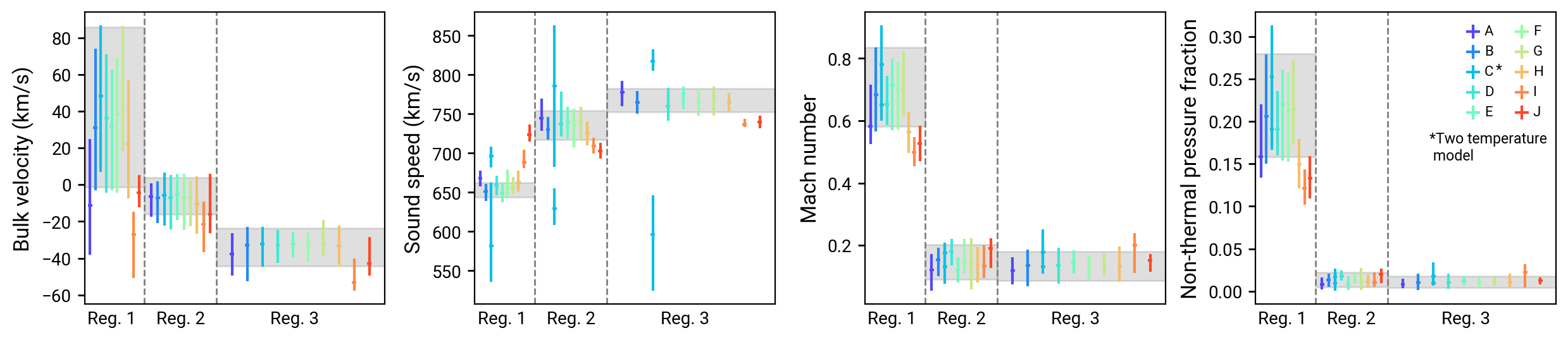}
    \caption{Distribution of the calculated properties of the ICM shown with 1$\sigma$ uncertainties. The four panels show (1) bulk velocity, (2) sound speed, (3) Mach number, (4) non-thermal pressure fraction. The gray shaded regions demarcate the statistical uncertainties from the default 4.2-7 keV band fit. The labels are the same as those in Fig.~\ref{fig:systematics}.}
    \label{fig:calculated}
\end{figure*}

The bulk velocity, shown in Figure~\ref{fig:calculated}, displays a mild trend from redshifted in the center, at +40\,km/s, to consistent with zero in the second region, to blueshifted in the northwest, at -30\,km/s, although with significant statistical and systematic uncertainties in region 1.

\subsection{Comparison to simulations and other methods}

Gas motions are expected to be driven on Mpc scales by mergers during the process of structure formation, and on smaller scales by the AGN, gas sloshing, galaxy motions, and instabilities \citep{ZuHone2013, Brueggen2015, Simionescu2019}. 
The velocity dispersion peak concentrated at the center of AGN activity, in the region of recently inflated bubbles and high pressure, followed by a sharp decline outside of this region, implies that the source of these motions is the AGN. This idea is supported by comparison with the velocities in sloshing-only simulations, which do not include the effects of other motions, such as AGN feedback. \cite{ZuHone2015} studied cold fronts due to gas sloshing in simulated cool-core clusters with initial conditions that correspond to the Virgo cluster. They found that, regardless of the simulation details (isotropic, anisotropic, varying viscosity), the velocity dispersion within $\sim 10$\,kpc of the center of the sloshing-only cluster is $\sim 20$\,km/s, suggesting that the high velocity dispersion measured with XRISM/Resolve is not driven by sloshing alone. 
Bulk velocities in the same region of these simulations are near zero, which is in agreement with our findings. This implies that sloshing could be primarily responsible for the bulk flows observed in the core.

Our findings can be further compared to indirect velocity measurements from surface brightness fluctuations \citep{Zhuravleva2014}. This method was performed on Chandra images of M87, excluding the central AGN and jet, and the arms. In that study, velocities within a 2’ radius circle are up to 110-140 km/s, which are consistent with the averaged velocity we measure with XRISM between regions 1 and 2. In a 2’-6’ annulus, the inferred velocities drop to 60-75 km/s, which are also within the range measured with XRISM in region 3. While a detailed comparison between the surface brightness fluctuation technique and XRISM’s direct velocity measurements is beyond the scope of this paper, the two methods yield similar velocities for the relaxed Virgo cluster -- consistent with the previously reported agreement in another relaxed cluster, Abell 2029 \citep{XRISM2025d,Heinrich2024}.

Constraints on the velocity dispersion in M87 were obtained with XMM-Newton RGS by \cite{Pinto2015}. In that work, the line broadening was used to infer upper limits on the velocity dispersion after accounting for spatial effects inherent to the slitless RGS design. Depending on the treatment of spatial broadening, the inferred 2$\sigma$ velocity broadening upper limit within 0.8\arcmin\,(3.7 kpc) is 210-470 km/s. Although the comparison is between regions of slightly different sizes (1\arcmin in this work and 0.8\arcmin\, in \cite{Pinto2015}), our XRISM measurement in the central region is consistent with the most conservative RGS upper limit.

The line shifts measured with XRISM can be compared to those measured in \cite{Gatuzz2022} with XMM-Newton. That work examined the redshift in a strip of the core directly with RGS data, and beyond $10$\,kpc with EPIC-pn data via background X-ray lines. They found a core bulk velocity consistent with the M87 optical velocity, in agreement with XRISM findings. Beyond 10 kpc, corresponding to region 3 in the XRISM analysis, \cite{Gatuzz2022} splits the data into a variety of regions. The ones most closely matching region 3 have line shifts either consistent with M87's optical redshift or blueshifted several hundred km/s, though with large uncertainties of similar magnitude. XRISM rules out such large bulk velocities but remains consistent within the XMM uncertainty range.

\subsection{Comparison to the dynamics of multiphase gas}

M87 hosts an extended network of optical H$\alpha$+[N II] filaments coincident with emission from a wide range of temperatures, including soft X-rays, far-UV C IV, far-IR C II, and submillimeter CO lines \citep[e.g.,][]{Ford1979,Sparks1993,Young2002,Sparks2004,Sparks2009,Werner2013,Simionescu2018}. This multiphase structure may arise from in-situ cooling of the hot ICM triggered by AGN feedback \citep{Gaspari2013a,Voit2015,McNamara2016}. Observations show that the kinematics of the cold and warm phases are remarkably consistent: CO and C II line velocities closely match those of the H$\alpha$ filaments \citep{Simionescu2018,Boselli2019}. In the case that the filaments do indeed condense out of the ambient X-ray plasma, their motions could reflect the velocity field of the surrounding hot gas if measured on the same scales.

The most detailed warm ($\sim10^4$ K) gas velocity maps come from Multi Unit Spectroscopic Explorer (MUSE) H$\alpha$ observations, which provide 0.2 arcsec spatial sampling and R = 3000 spectral resolution over a 1\arcmin x 1\arcmin\ field -- roughly matched to the extent of our Region 1 \citep{Sarzi2018,Boselli2019}. 
The H$\alpha$ filaments show disordered motions with $\sim$100 km/s line widths and point-to-point velocity differences reaching 700–800 km/s \citep{Boselli2019}. From these data, \citet{Li2020} derived a velocity structure function (VSF) peaking at an average fluctuation $<|\delta v|>$ $\sim150$~km/s on $\sim2$~kpc (0.4 arcmin) scales. On much smaller scales, ALMA resolves a single $\sim$ 150 pc ($\sim$ 2 arcsec) CO-emitting clump located $\sim 40$ arcsec southeast of the nucleus, with a  velocity dispersion of $27\pm3$ km/s \citep{Simionescu2018}, consistent with the 30 km/s MUSE VSF prediction at $\sim$0.1 kpc \citep{Li2020}. Assuming that the CO clump and H$\alpha$ filaments occupy similar volumes along the line of sight, the available data suggest that the different cool phases form a coherent velocity field across nearly two orders of magnitude in physical scale.

Given this apparent coherence among the cool phases, the key question we now explore is whether the X-ray–emitting gas, which traces a distinct and more volume-filling phase of the ICM, follows the same velocity–scale relation. The XRISM measurement of a velocity dispersion $\sigma_v=262^{+45.0}_{-38.4}$ km/s in Region 1 exceeds the $\sim150$~km/s reported for the H$\alpha$ filaments in \cite{Li2020}. Several factors could account for this difference. XRISM captures motions of the diffuse, volume-filling plasma, whereas filaments occupy a small fraction of that volume. Some of the observed X-ray line width may also reflect laminar outflows and the expansion of a shocked shell, rather than the small-scale turbulence that filaments may trace. Alternatively, the dynamics of the line-emitting nebulae may be partially or fully decoupled from those of the hot phase, as suggested by XRISM observations of the Centaurus BCG, where the ICM is blue shifted relative to the H$\alpha$ emission \citep{XRISM2025c}. The hypothesis that the velocity structure of line emitting nebulae traces that of their ambient X-ray plasma should therefore be reexamined carefully with future observations. 

\subsection{Energetics, Heating, Cooling} \label{subsec:heating}

Although the increased velocity dispersion at the cluster center is likely driven by AGN feedback, the observed gas motions may arise from a combination of physical processes. Potential sources of motion include directed outflows from the AGN, the inflation or buoyant rise of radio bubbles, shocks, and turbulence. Using the current observational constraints alone, it is difficult to distinguish whether the line-of-sight velocity components trace directed outflows or random turbulent motions.  As a result, assumptions must be made about the nature of motions, which can influence our estimates of energy injection by the AGN.

Assuming that the observed velocity dispersion results entirely from an isotropic turbulent velocity field, and that the ICM behaves as an ideal monatomic gas, we can estimate the three-dimensional Mach number. For this, we first calculate the sound speed as 
\begin{equation}
c_s = \sqrt{\frac{\gamma k_\mathrm{B} T}{\mu m_p}},
\end{equation}
where $\gamma = 5/3$ is the adiabatic index for monatomic gas, $k_B$ is the Boltzmann constant, $T$ is the measured gas temperature, $\mu$ is the mean molecular weight, and $m_p$ is the proton mass. The sound speed, like the temperature, is lowest in the central region ($c_s \sim 660$\,km/s), increases between regions 1 and 2, and is consistent between regions 2 and 3 ($c_s \sim 750$\,km/s).

The three-dimensional effective Mach number, accounting only for velocity dispersion, is defined as 
\begin{equation}\label{eqn:mach}
M_{\mathrm{3D,eff}} = \frac{\sqrt{3}\sigma_v}{c_s},
\end{equation}
where $\sigma_v$ is the line-of-sight turbulent velocity dispersion. Under these assumptions, the Mach number in the central region is in the transonic regime ($M = 0.69^{+0.14}_{-0.11}$). The Mach number then drops steeply from region 1 to region 2 and remains consistent from region 2 to 3, at a value of $\sim 0.15$.

Using the estimate of $M_{\mathrm{3D,eff}}$, we can then estimate the ratio of non-thermal pressure to total pressure using
\begin{equation}
\frac{P_{NT}}{P_{tot}} = \frac{M_{\mathrm{3D,eff}}^2}{M_{\mathrm{3D,eff}}^2 + \frac{3}{\gamma}}.
\end{equation}
This presumes that the non-thermal pressure arises entirely from kinetic motions and ignores potential contributions from magnetic fields or cosmic rays. The non-thermal pressure fraction reaches $\sim 20\%$ in the central region, and shows a steep drop between regions 1 and 2, similar to the Mach number. The largest uncertainties are in region 1. 

For all calculated values (shown in Figure \ref{fig:calculated}), uncertainties from the fits were propagated through the equations, including covariance terms, to obtain the errors.

\subsubsection{Propagating weak shocks and sound waves}\label{sect_sound}

Deep Chandra observations of M87 reveal an overpressurized shell inside the central radio cocoon (within region 1), and a circular shock front at approximately 13 kpc (within our region 3). The inner shell is thought to be associated with the ongoing AGN outburst, while the shock front likely traces a previous outburst episode \citep{Forman2017}. If the velocity field is dominated by propagating weak shocks and sound waves, we may expect a sharply peaked velocity dispersion towards the cluster center. The velocity dispersion profile shown in Figure 6 indeed reveals a sharp jump in the velocity dispersion profile between regions 1 and 2, which corresponds in size and may be associated with the expansion of the inner shell. 

\begin{figure}
    \centering
    \includegraphics[width=0.8\linewidth]{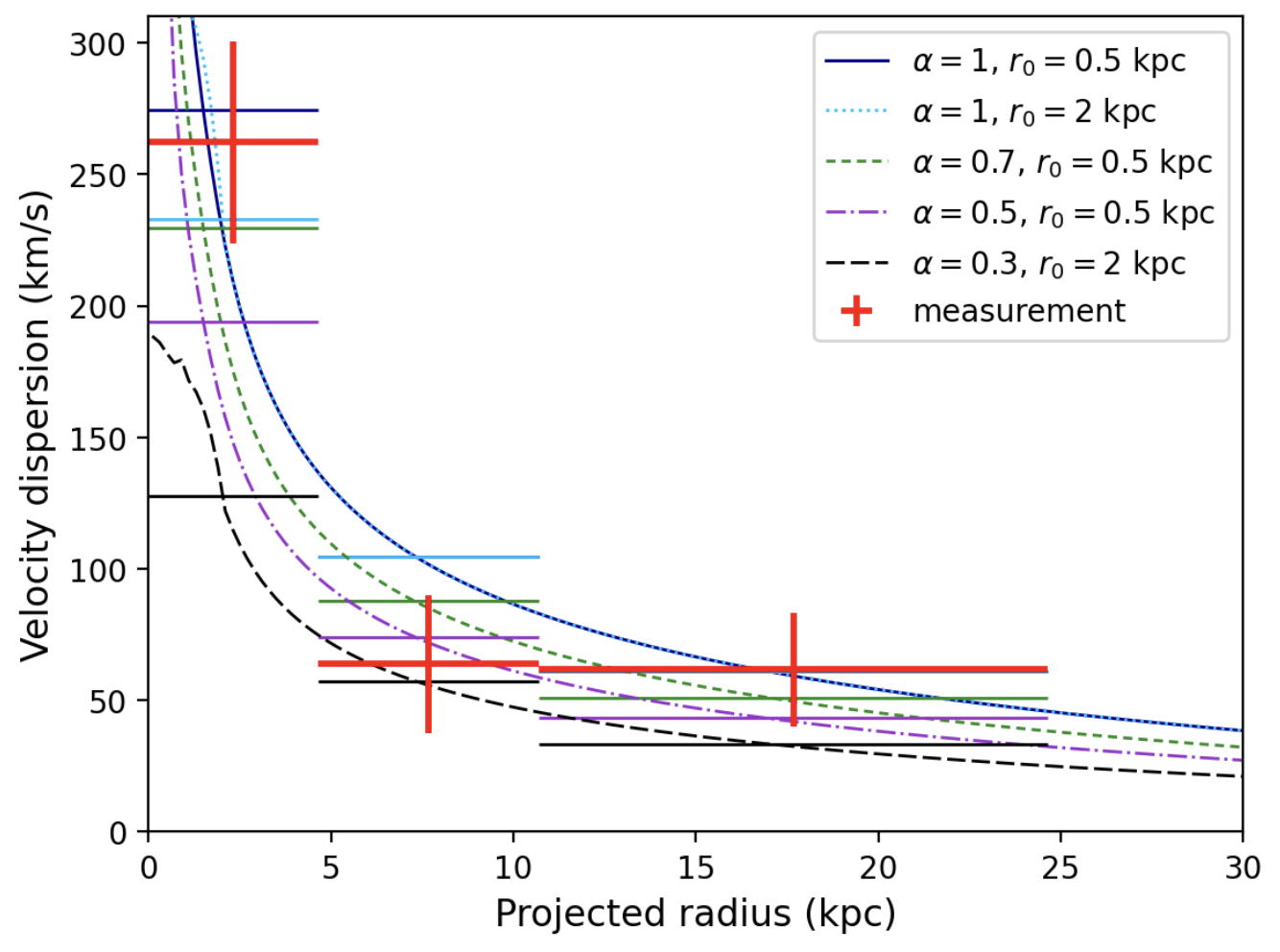}
    \caption{Modeled radial profiles of velocity dispersion  produced by propagating sound waves, assuming the waves  originate at cluster radius $r_{0}$ and offset a fraction ($\alpha$) of cooling losses everywhere within the 100 kpc radius. Both smooth and binned profiles are shown. Red points show the velocity dispersion measured in this work.}
    \label{fig:sw}
\end{figure}

To check this quantitatively, we applied a simple, spherically-symmetric model of propagating sound waves discussed by \cite{Fabian2017}.  According to this model, the energy flux of sound waves, $F_{\rm sw}$, across the radius $r$ is conserved as $F_{\rm sw} = 4\pi r^2 \rho (r) v_{\rm s} (r)^2 c_{\rm s}$, where $v_{\rm s} (r)$ is the wave velocity amplitude and $\rho(r)$ is the gas mass density. Assuming that sound waves balance a fraction of cooling losses between the radii $r_0$ (e.g., the size of the inner bubble) and $r_{\rm max}$ (e.g., the size of the cool core), we can relate $F_{\rm sw}$ to radiative cooling losses $Q_{\rm cool}$ as $d F_{\rm sw}(r)/d r=\alpha\cdot 4\pi r^2 Q_{\rm cool}$. We compute the cooling rate as $Q_{\rm cool}=n_e n_H\Lambda$, with $\Lambda = 1.84 \times 10^{-23}$ erg cm$^3$/s adopted from the recent calculations by \cite{Stofanova2021} and estimated at Solar abundance and a temperature of 2 keV (i.e. neglecting a small dependence of $\Lambda$ on the exact gas temperature), and $n_e=1.2n_H$ for a fully ionised plasma with Solar abundance. For the electron density profile, we have reanalyzed the Chandra data as presented in \cite{Russell2015}, with updated atomic line libraries and the Solar metal abundance reference table of \cite{Lodders2009}. We combined this Chandra-based profile with the $eROSITA$ measurements at larger radii \citep{McCall2024} and approximated the whole profile with a double $\beta-$model, namely 

\begin{equation}\label{eqn:density}
    n_{\rm e} = 0.42 \left[1 + \left(\frac{r}{0.51}\right)^2\right]^{-0.57}+ 0.008\left[1 + \left(\frac{r}{24.6}\right)^ 2\right] ^{-1.13},
\end{equation}
 where $r$ is measured in kpc and $n_e$ is in cm$^{-3}$. The data and best-fit model are shown in Fig.~\ref{fig:density}. We further assume that the inner bubbles are empty of gas, that $F_{\rm sw}=0$ at $r_{\rm max}$, and that the factor $\alpha$, which represents the fraction of cooling losses that are compensated by heating due to sound waves, is constant as a function of radius. We caution the reader that the latter assumption in particular may not be correct; however, lacking a detailed model of the dissipation of sound wave energy into the ICM, such simplifications are necessary to gain a zeroth-order understanding of this heating model.

Figure \ref{fig:sw} presents a selection of models with the free parameters $\alpha$, $r_{0}$, and $r_{\rm max}$. If sound waves are assumed to fully offset cooling losses ($\alpha = 1$, solid and dotted blue curves in Fig. \ref{fig:sw}), the size of the bubble producing these waves is required to be smaller than a few kpcs to roughly reproduce the observed sharp decline in velocity dispersion. However, the transition between regions 1 and 2 remains difficult to describe, as the theoretical profiles appear shallower than the observed jump in the velocity dispersion. Moreover, $\alpha = 1$ is at odds with previous studies, which suggest that only about 20–30\% of the outburst energy is carried by shocks \citep{Forman2017}. Models with reduced $\alpha$ (shown as dashed and dot-dashed curves in Fig. \ref{fig:sw}) struggle to reproduce the steep observed drop in velocity dispersion unless the bubble size is comparable to or smaller than 0.5 kpc, which is smaller than the typical observed bubble size. The same is true when measurements other than those resulting from the default model (i.e., from the broad band 3-11 keV fits) are considered. These tensions suggest that the measured velocity dispersion in the central region likely reflects contributions from both the expansion velocity of a shock-driven shell and additional random gas motions. Resolving these components requires tailored numerical simulations of AGN feedback in M87, which is beyond the scope of this paper. It is also worth noting that the sound wave model is likely an oversimplification in this case, as the observed velocity profile probably reflects a superposition of multiple outburst episodes.

\subsubsection{Dissipation of random gas motions}

As mentioned above, weak shocks and sound waves alone appear unlikely to explain the peaked velocity dispersion profile measured in M87 by XRISM Resolve. Consequently, we consider a scenario where all of the observed velocity broadening is due to random motions.  

Following an approach similar to \cite{Rose2025}, we start by comparing the kinetic energy of the atmosphere, evaluated as $3/2 M_{\rm atm}(<r) \sigma^2$, to the atmospheric X-ray luminosity, given an assumed dissipation timescale, to determine whether turbulent heating is in principle capable of offsetting radiative cooling. The gas mass was estimated using the observed electron density from Chandra (see Figure~\ref{fig:density}). 
We then assume that the velocity dispersion measured by XRISM is entirely attributed to well-developed turbulence, that the dissipation time scale $\tau$ corresponds to the 12 Myr AGN duty cycle (inferred based on the age of the outer shock by \cite{Forman2017}), and that all turbulent kinetic energy goes to heat. 

In this ideal case, for central region 1, which has $\sigma\sim262$ km/s, the kinetic power
\begin{equation}\label{eqn:turbpower}
P_{\rm turb}=3/2 M_{\rm atm}(<r) \sigma_i^2 / \tau
\end{equation}
 is $6\times10^{42}$ erg/s when we take the measured velocity dispersion $\sigma$ as $\sigma_i$, the velocity dispersion integrated over the entire range of scales from injection to dissipation. This assumption likely means that we are underestimating $\sigma_i$, since we do not necessarily measure the full cascade. The cooling luminosity is roughly a factor of 5 lower than the turbulent power, at $1.1 \times10^{42}$ erg/s.
 Assuming a dissipation time even shorter than 12 Myr, such as, for instance, the age of the inner shell, currently estimated at 2 Myr, only enhances the mismatch between the kinetic power and cooling rate, as well as between the inferred injection scale and bubble sizes.
 This supports our conclusion from the previous subsection, that the velocities in this central region must be a superposition of turbulent and non-turbulent motions due to shocks, sound waves, or to unresolved bulk motion of material displaced along the line of sight by the jet.

\begin{figure*}
    \centering
    \includegraphics[width=0.4\linewidth]{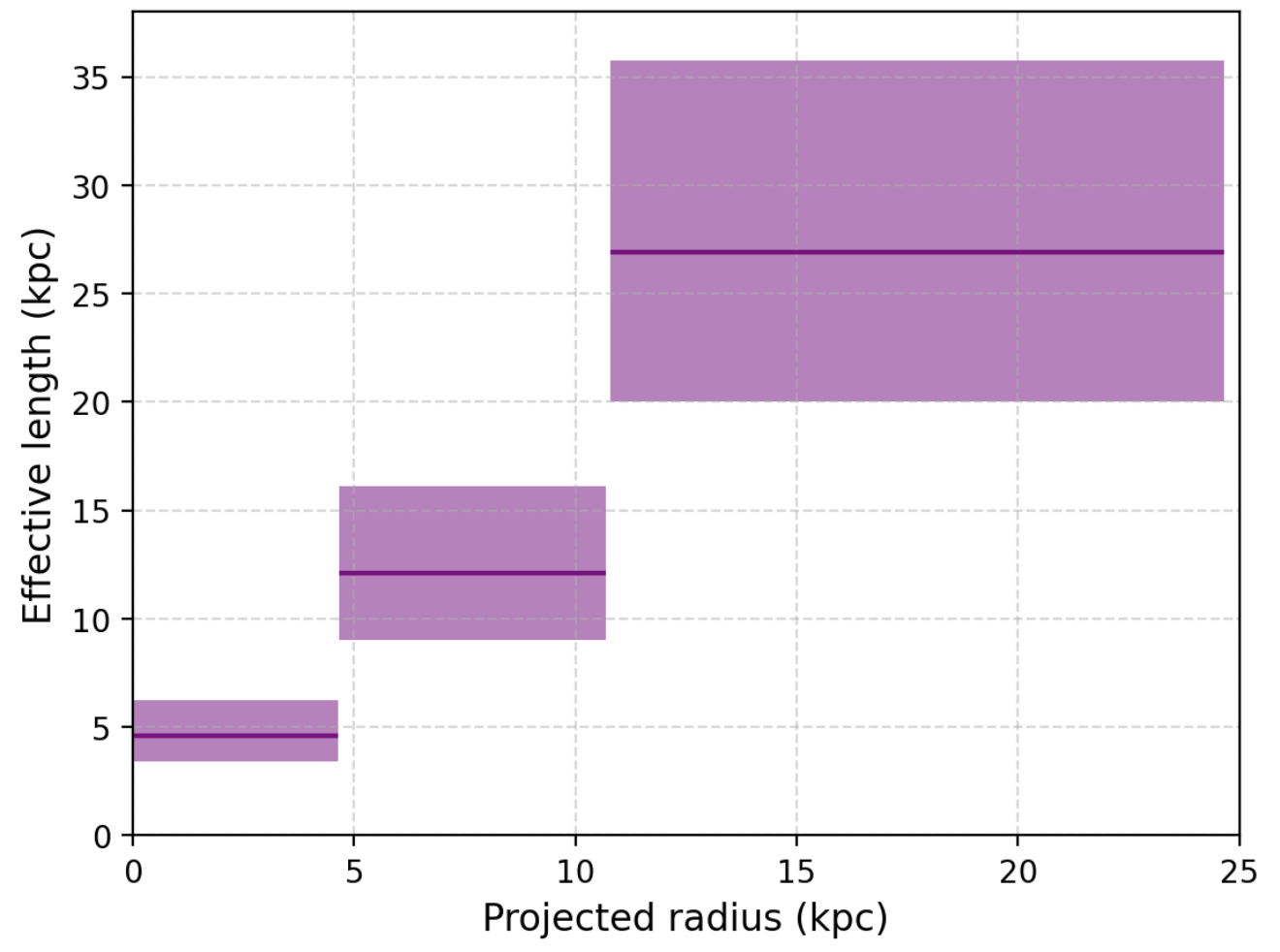}
    \includegraphics[width=0.4\linewidth]{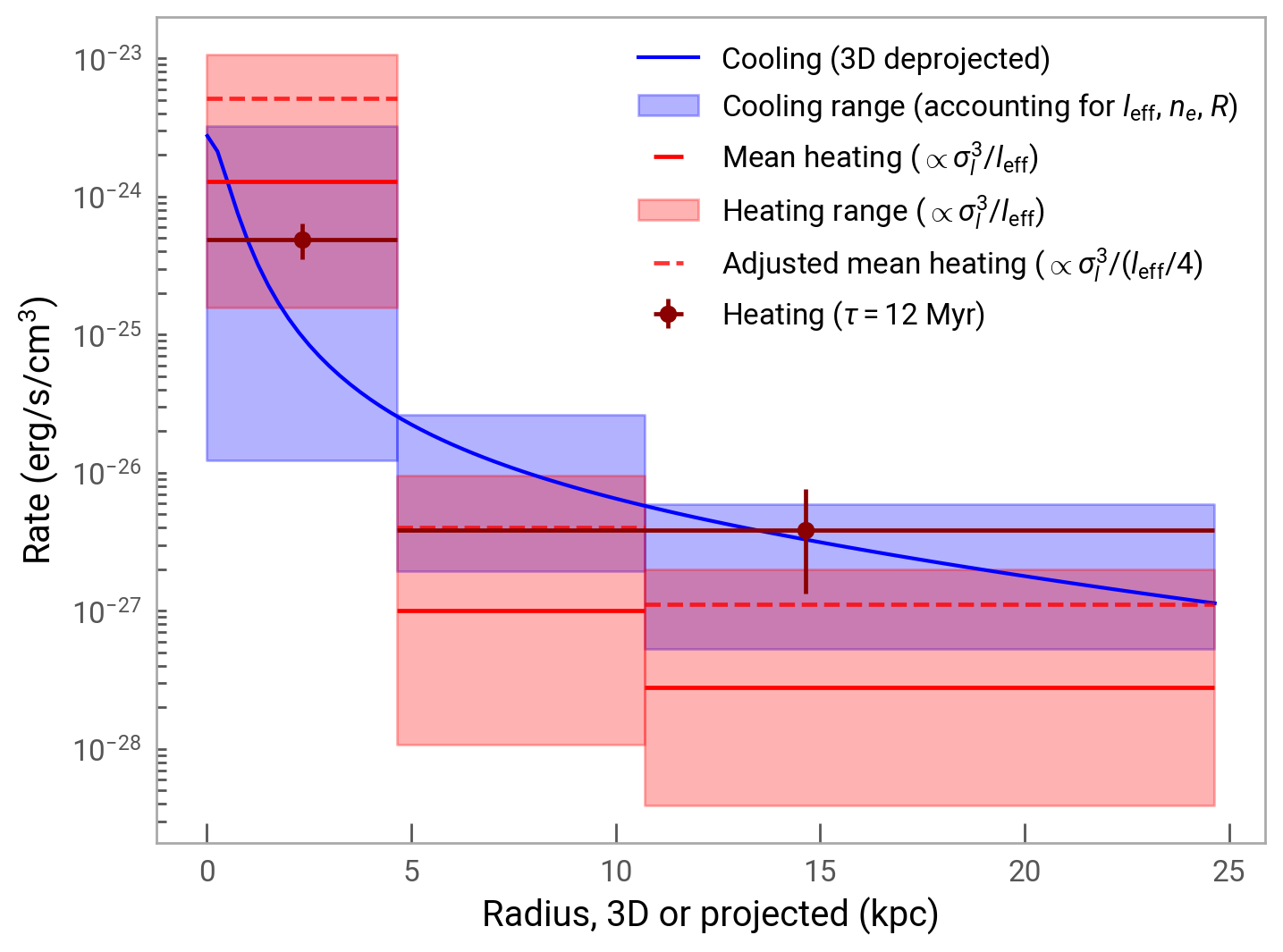}
    \caption{
    Left: Effective length scale, $l_{\rm eff}$, of the Virgo cluster averaged within the projected radial bins considered in this work. The solid lines show the effective length defined as the region size where 50\% of the flux is collected, while the shaded region indicates the scales associated with 40-60 \% of the flux contribution. 
    Right: Cooling (blue) and estimated heating (red) rates in the core of the Virgo Cluster. Due to density variations within the large spatial regions studied with XRISM, as well as uncertainties in the physical spatial scales associated with the measured velocity dispersion, there are large uncertainties on both rates. 
    In detail, the blue line shows the cooling rate as calculated from the deprojected density profile described in Eq.~\ref{eqn:density}. The blue boxes show the cooling rate binned to the regions observed with XRISM Resolve, with uncertainties due to the variation of the cooling within each bin and due to the range of effective lengths contributing to the measurement (left figure). Heating is calculated via two approaches; the solid red line and accompanying red boxes show the result of Eq.~\ref{eqn:heatturb} using the nominal values and with uncertainties due to velocity measurements and $l_{\rm eff}$, while the dark red points are calculated from Eq.~\ref{eqn:turbpower} using an assumed dissipation timescale. The dashed red lines use the same approach as the solid red lines, but with a scale of a factor 4 smaller to show the effect of assuming that the scale $l_\mathrm{eff}$ associated with $\sigma$ is smaller than the conservative upper limit.
}
    \label{fig:ch_leff}
\end{figure*}

On the other hand, in regions 2 and 3, where the velocity dispersion drops to 60 km/s, the X-ray luminosity and atmospheric kinetic energy are roughly matched, with 
$P_{\rm turb}\sim 7\times 10^{42}~\rm erg~s^{-1}$ and $L_{\rm x}\sim 5.3\times 10^{42}~\rm erg~s^{-1}$. Therefore, under these ideal assumptions, turbulent dissipation would be energetically capable of approximately balancing radiative cooling. However, a turbulent dissipation timescale of $\tau = 12~\rm Myr$ implies a turbulence injection scale $l_i \simeq \tau \sigma_i \simeq 0.7 \rm ~kpc$. This scale is considerably smaller than the $3-5$ kpc bubble sizes in M87 (Forman et al. 2017). This discrepancy could be due to our assumption that $\sigma = \sigma_i$ when $\sigma$ is only a measure of motions on scales below 5 kpc, leading to an underestimation of $\sigma_i$.
If, however, we assume an injection scale lying between $3-5 \rm ~kpc$, approximately the size of the observed bubbles, we can rewrite the power as:
$P_{\rm turb}\sim3/2 M_{\rm atm}(<r) \sigma_i^3 / l_i$.  We then find a turbulent dissipation power of $1-2\times 10^{42} ~\rm erg ~s^{-1}$, which lies below the cooling power, but within the expected order-of-magnitude uncertainty range.
These scales lie far below XRISM’s spatial resolution. Therefore, the relevant $\sigma$ on these scales cannot be directly probed by this observation. 

We find that (under the same assumptions) the sum of the kinetic energy enclosed within the radius of the outer shock corresponds to roughly half of the $5\times 10^{57}\rm~ erg$ outburst energy reported by \cite{Forman2017}, indicating that the AGN is energetically able to produce turbulence at the observed level. The calculations presented above do not make any assumptions about the shape of the turbulent power spectrum or the atmospheric fluid properties other than that turbulence can be represented as an isotropic, Gaussian-random field. They do, however, make assumptions about the timescale of dissipation and the role of $\sigma$ as $\sigma_i$.

The dissipation time scale is not well known, and the assumption that it matches the AGN duty cycle at all radii is somewhat arbitrary. 
In the absence of a measurable dissipation time scale or injection scale, an alternative way to estimate the rate of turbulent dissipation is described by \citet{Kolmogorov1941}. If, namely, we can probe the turbulent power spectrum describing the dependence of the velocity amplitude $\sigma_l$ as a function of spatial scale $l$, and assuming $l$ is in the inertial range of the turbulent cascade, the heating rate can be calculated as 
\begin{equation}\label{eqn:heatturb}
Q_{\rm heat} = C_0 \rho(r) \sigma_l^3/l. 
\end{equation} 
Here, $C_0= 3^{3/2} 2\pi / (2 C_K)^{3/2}$, and $C_K$ is the Kolmogorov constant, which varies in a relatively narrow range for known, physically motivated examples of turbulent power spectra (\cite{Sreenivasan1995,Kaneda2003}), and which we take as $C_K\approx1.65$. The advantage of this approach is that there is a clearly defined (from a theoretical point of view) correspondence between $\sigma_l$ and a spatial scale $l$, without any assumptions on injection or integrated scales that are often difficult to determine or even define unambiguously.

We note in passing that in the Coma Cluster, \cite{XRISM2025b} report a steep turbulent power spectrum that is difficult to reconcile with theoretical expectations. This may raise questions as to whether `simple' physical models such as the one used in Eqn. \ref{eqn:heatturb} can be easily applied for the ICM. 
An additional difficulty in using this equation arises from the fact that the velocity dispersion $\sigma$ measured by XRISM does not probe any single spatial scale. However, for a physically-motivated picture where the energy is injected on large scales and cascades down to small scales, it is not unreasonable to assume that the measured $\sigma$ is dominated by the largest scale within the region where most of the emission originates. 

Since the X-ray emissivity (and gas density) is typically peaked towards the cluster centers (including Virgo), the portion of the LOS from which most of the flux originates is shortest near the cluster core and increases with projected radius $R$ \citep{Zhuravleva2012}. This means that velocities measured in the center and at some projected distance could be associated with very different length scales. To this end, it is useful to define an effective length $l_{\rm eff}$, following \cite{Zhuravleva2012}, as the distance along the LOS that accounts for 50\% of the total X-ray flux at a given projected radius. Using the deprojected density profile, we estimated $l_{\rm eff}$ to be $\sim 4-6$, $12-16$, and $27-36$ kpc in the radial bins 1-3, respectively. This result is displayed in Figure~\ref{fig:ch_leff}. The associated scatter was defined by the regions contributing 40\% and 60\% of the flux. This sets the physical scale along the LOS that contributes with the highest weight towards our $\sigma$ measurements in each region. Motions on scales larger than $l_{\rm eff}$ in essence are heavily down-weighted due to the lower flux of that gas, and do not contribute significantly to our measurements.

Given the discussed assumptions and calculated $l_{\rm eff}$, we can calculate the heating rate (or at least its lower limit, given $l_{\rm eff}$ is the upper limit on contributing scales) due to the dissipation of measured motion using Eq.~\ref{eqn:heatturb}.
The height of each red region reflects the uncertainties on $\sigma$ and $l_{\rm eff}$ in each radial bin and on $\rho$, both in radial bins and within $l_{\rm eff}$. It is interesting to compare the heating rate with the corresponding cooling rate (blue regions in Figure \ref{fig:ch_leff}). The height of each blue region reflects the cooling rate variations within the observed projected radii and along the LOS up to $l_{\rm eff}$.  

Figure \ref{fig:ch_leff} shows several noteworthy features. 
Within the admittedly large uncertainties, there is an approximate balance between radiative cooling (calculated as described in Section~\ref{sect_sound}) and turbulent heating.
However, as argued previously, it is likely that the observed motions are due in part to the expanding shell around a recent AGN outburst, rather than exclusively to turbulence.
In the case where we impose a precise match between the mean turbulent heating and cooling rates, the portion of $\sigma$ attributed to the expanding shell would be $160$\,km/s, leaving $210$\,km/s driven by random motions.
Future numerical simulations of AGN feedback in Virgo will be crucial for disentangling the contributions of bulk and random motions in the inner 5 kpc, and to clarify their respective roles in offsetting radiative cooling.

For regions 2 and 3, instead, the implied mean heating rate computed under the above assumptions falls somewhat below the mean cooling rate of the atmosphere. However, both are consistent within the uncertainties related primarily to $l_{\rm eff}$ and $\rho$, but also to the statistical errors on $\sigma$ (see Table \ref{tab:results}). The difference between mean values would be mitigated if the characteristic scales associated with the measured velocities were smaller than $l_{\rm eff}$.
This would be the case if the turbulence cascade peaks at $l_{\rm inj} < l_{\rm eff}$; then, eddies on scales larger than $l_{\rm inj}$ would not contribute significantly to $\sigma$ and the measured $\sigma$ would correspond to the $\sigma$ at $l_{\rm inj}$. 
Keeping $\sigma$ as observed, but assuming characteristic scales are a factor of four lower, which are reasonable given the sizes of structures observed within these regions (e.g., shocks, bubbles), the resulting heating rate would well balance the cooling rate (dashed line in Fig. \ref{fig:ch_leff}). 
Interestingly, velocity power spectra of M87 measured through X-ray surface brightness fluctuations show velocities comparable to those measured here with XRISM if, indeed, the scales associated with these motions are within the $3-10$ kpc range \citep{Zhuravleva2014}, consistent with $l_\mathrm{eff}/4$. 

Given the measured velocity dispersion of $\sim 260$ km/s on scales $\lesssim 5$ kpc and of $\sim 60$ km/s on scales $\lesssim 25$ kpc (averaging over regions 2-3), we can conclude that a single physically-motivated velocity cascade (i.e., where energy is pumped in on large scales and cascades down to small scales) cannot explain the velocity field in M87. Rather, at least two drivers are required: one operating on small scales in the innermost regions and another in regions 2-3. The same behavior was found in the Perseus Cluster with XRISM, where more than one driver was found to be required to explain the motions within 60 kpc and at larger scales \citep{XRISM2025e}. However, given that a fraction of kinetic energy in the innermost region in M87 is likely associated with an expanding shell around the central cocoon rather than random motions, this indication should be further explored with tailored simulations of feedback in M87.

While the two approaches to estimating turbulent heating---via total kinetic energy, or via the Kolmogorov cascade framework---rely on distinct assumptions about the spatial and temporal scales associated with turbulence, they yield broadly consistent heating rates. 
XRISM has allowed us to quantify the kinetic energy of the Virgo ICM with unprecedented precision.
Nonetheless, given the unknowns surrounding the dissipation timescale, the injection scales, and the contribution of non-turbulent motions to the observed velocity dispersion, considerable uncertainty remains in quantifying the exact role of turbulence in the energy balance of the ICM.

\begin{figure*}
    \centering
    \includegraphics[width=0.46\linewidth]{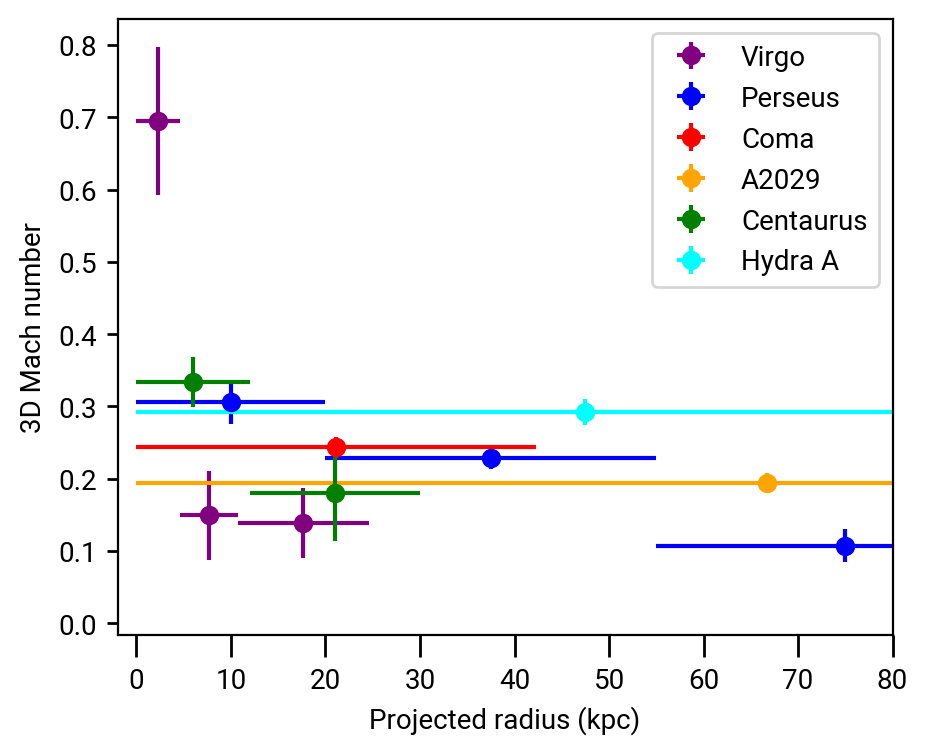}
    \includegraphics[width=0.46\linewidth]{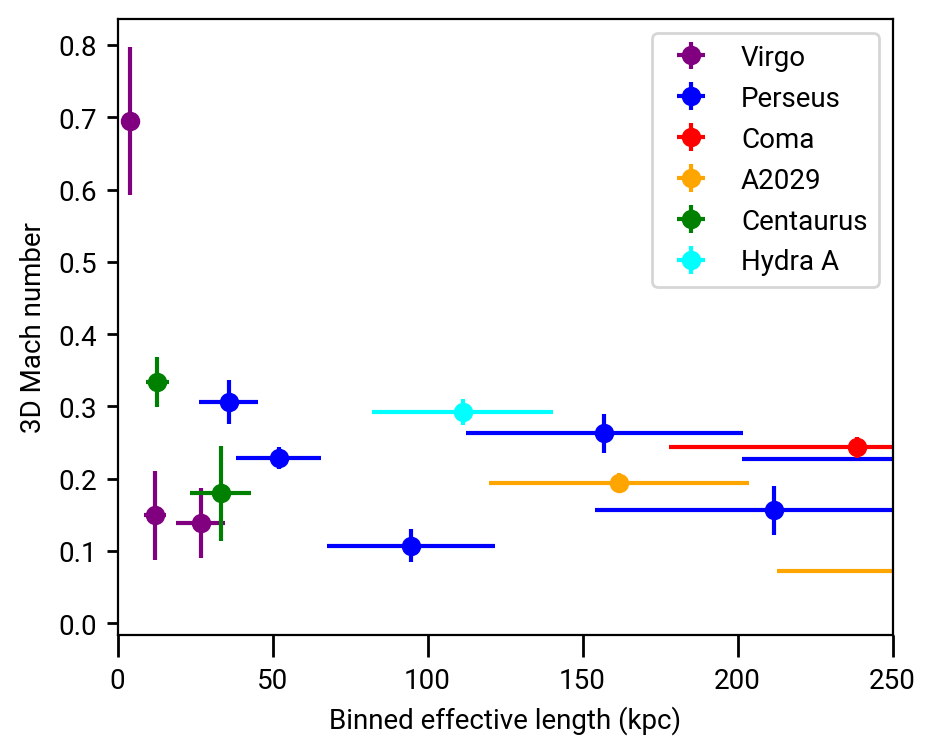}
    \caption{3D Mach number as a function of (left) projected radius and (right) effective length for a selection of clusters observed with XRISM Resolve.}
    \label{fig:comparison}
\end{figure*}

\subsection{Comparison with other clusters}
 \label{sec:comparison}   

 A selection of other clusters, both with and without AGN feedback, has also been observed with XRISM/Resolve to date. In this work we consider for comparison the feedback-active systems Perseus \citep{XRISM2025e}, Centaurus \citep{XRISM2025c}, and Hydra A \citep{Rose2025}; the relaxed cluster Abell 2029 \citep{XRISM2025a}\citep{XRISM2025d}; and the merger Coma \citep{XRISM2025b}. Comparing the central velocity dispersion in M87 to the velocity dispersions measured in these clusters, M87 boasts the highest value ($\sigma_v = 262$\,km/s), larger even than the merging Coma Cluster ($\sigma_v = 208 \pm 12$ km/s). However, a direct comparison between the velocity dispersions measured in these clusters ignores that these measurements arise from different scales and from regions with different sound speeds. For a fairer comparison of the kinematic behavior of these clusters, we consider the 3D Mach number as a function of radius.

We calculate the 3D Mach numbers using only the velocity dispersion from each source, following Eq.~\ref{eqn:mach}. The effective lengths as a function of projected radius are calculated as introduced in subsection~\ref{subsec:heating}, using $\beta$-model values taken from literature for each cluster. We then take the average effective length within a given projected radius, with the projected radius depending on the size of the XRISM/Resolve subregion in each case (i.e. the width of the bins in the left plot of Fig.~\ref{fig:comparison}). The uncertainties on the binned effective length are the scales associated with 40\% and 60\% of the flux contribution. Fig.~\ref{fig:comparison} shows a comparison of the 3D Mach numbers of observed clusters as a function of projected radius (left) and effective length (right). As evidenced by this figure, the observed Mach number in the center of M87, $\sim 0.6-0.8$, at scales of less than 7\,kpc, is a spectacular kinematic feature that is not only significantly higher than in any other cluster, but also originates from smaller scales, both projected and along the line-of-sight. Although other clusters with multiple data points also display a generally decreasing trend at scales below 80\,kpc, the drop from the innermost 5-7\,kpc of M87 to the next radial bin is also notably steep.

\section{Conclusions}
In this study, we used XRISM/Resolve to investigate the radial velocity structure of the Virgo Cluster in a direction unaffected by the large-scale arms.
We divided two $3\arcmin \times 3\arcmin$ pointings into three sub-regions and found a peaked central velocity dispersion of $\sigma_v=262^{+45}_{-38}$ km/s around the AGN and jet, with a steep decline to $\sim 60$\,km/s beyond 5\,kpc. The bulk velocity, on the other hand, was found to be near zero in all sub-regions, with a mild trend from redshifted to blueshifted with increasing radius. These findings are robust to systematic uncertainties.

The peaked central velocity dispersion corresponds spatially with known AGN-driven structures, suggesting that AGN feedback is the primary source of motions in this region. Assuming isotropic turbulence, we estimate a central Mach number of $M \sim 0.69$, corresponding to a non-thermal pressure fraction of $21\%$. The measured velocity dispersions are consistent with indirect estimates from surface brightness fluctuations.

Simple spherical sound wave models can reproduce the steep central velocity dispersion gradient only under extreme assumptions (i.e., very small bubble sizes) that are inconsistent with observations, suggesting that the velocity field is a combination of shock-driven expansion and random motions. 

We estimate the turbulent heating using two different methods, each with its own assumptions and limitations. 
While our results allow for a contribution of AGN-driven motions to heating the Virgo Cluster’s core, substantial modeling uncertainties preclude a definitive assessment of their role.

Comparisons with the kinematics of multiphase gas 
suggest a discrepancy between the hot ICM and cooler filaments in the innermost region. This could be due to a dynamical decoupling of the various gas phases, the different volume-filling factors at play, or to the fact that cooler filaments trace only the turbulent velocity, while the X-ray velocity dispersion includes the expansion of the inner shock.

When compared to other clusters observed with XRISM/Resolve, M87 stands out as the most kinematically disturbed in the core. Its central velocity dispersion exceeds that even of the merging Coma Cluster, and its 3D Mach number---0.6-0.8 at sub-7 kpc scales---is uniquely high, both in absolute terms and relative to the size of the region probed.

Future works will present the kinematic results of the XRISM/Resolve observations of the M87 arms to the east and southwest, and the temperature and abundance findings for all observations. Re-observing with XRISM's gate valve open would allow access to lower-energy spectral lines, enabling the study of cooler gas phases in the multiphase arms of M87. In the longer term, ESA's upcoming \textit{NewAthena} mission, with its superior spectral and spatial resolution, will be instrumental in resolving the small-scale velocity structure and mapping turbulence across the ICM, building on the insights into AGN feedback in cluster environments provided by XRISM.

\begin{acknowledgments}
The results presented above are made possible by over three decades of work by the team of scientists and engineers who created a microcalorimeter array for X-rays and overcame enormous setbacks. We gratefully acknowledge the entire XRISM team's effort to build, launch, calibrate, and operate this observatory.
Part of this work was supported by the U.S.\ Department of Energy by Lawrence Livermore National Laboratory under Contract DE-AC52-07NA27344, and by 
NASA under contracts 80GSFC21M0002 and 80GSFC24M0006 and grants 80NSSC20K0733, 80NSSC18K0978, 80NSSC20K0883, 80NSSC20K0737, 80NSSC24K0678, 80NSSC18K1684, 80NSSC23K0650, and 80NNSC22K1922.
Support was provided by JSPS KAKENHI grant numbers JP23H00121, JP22H00158, JP22H01268, JP22K03624, JP23H04899, JP21K13963, JP24K00638, JP24K17105, JP21K13958, JP21H01095, JP23K20850, JP24H00253, JP21K03615, JP24K00677, JP20K14491, JP23H00151, JP19K21884, JP20H01947, JP20KK0071, JP23K20239, JP24K00672, JP24K17104, JP24K17093, JP20K04009, JP21H04493, JP20H01946, JP23K13154, JP19K14762, JP20H05857, JP25K23398, and JP23K03459, the JSPS Core-to-Core Program, JPJSCCA20220002, and the Strategic Research Center of Saitama University.
LC acknowledges support from NSF award 2205918. 
CD acknowledges support from STFC through grant ST/T000244/1. 
LG acknowledges support from Canadian Space Agency grant 18XARMSTMA.
NO acknowledges partial support by the Organization for the Promotion of Gender Equality at Nara Women's University. 
MS acknowledges support by the RIKEN Pioneering Project Evolution of Matter in the Universe (r-EMU) and Rikkyo University Special Fund for Research (Rikkyo SFR). 
AT acnowledges support from the Kagoshima University postdoctoral research program (KU-DREAM). 
SU acknowledges support by Program for Forming Japan's Peak Research Universities (J-PEAKS).
SY acknowledges support by the RIKEN SPDR Program. 
IZ acknowledges partial support from the Alfred P.\ Sloan Foundation through the Sloan Research Fellowship.
\end{acknowledgments}

\begin{contribution}
H. McCall performed the XRISM analysis, wrote much of the manuscript, and created most of the figures.
A. Simionescu devised the data acquisition plan, coordinated the project, provided conceptual input for the discussion section, and contributed to the writing. 
C. Kilbourne contributed technical knowledge related to instrumentation and data interpretation and edited the manuscript.
H. Russell contributed expertise on the AGN, jet, and LMXBs, performed the Chandra analysis and wrote the corresponding sections.
A. T\"{u}mer provided scripts for data reduction, performed the NuSTAR analysis and wrote the corresponding sections.
I. Zhuravleva supervised HM, provided conceptual input, contributed to the writing, created several figures, and edited the manuscript.
B. McNamara provided conceptual input for the discussion section and contributed to the writing.
N. Dizdar, D. Eckert, Y. Ichinohe, D. Ito, M. Loewenstein, J. Martin, A. Ogorzalek, HR, K. Sato, AS, AT, and IZ independently analyzed the data.
DE, L. Gu, E. Hodges-Kluck, CK, M. Leutenegger, ML, BM, F. Mernier, E.D. Miller, K. Nakazawa, AO, KS, AS, A. Szymkowiak, AT, IZ, DI, JM, HM, and HR discussed the results and provided feedback as part of the target team.
J. ZuHone helped improve the manuscript.
The scientific objectives of XRISM were shaped over a period of seven years by the XRISM Science Team, whose members are all co-authors of this paper. The development of all instruments was a collaborative effort by the team. This manuscript underwent a thorough internal review across the collaboration, and all authors have reviewed and approved its final version.
\end{contribution}

\section*{Data Availability}
This article employs a list of Chandra datasets, obtained by the Chandra X-ray Observatory, contained in~\dataset[doi: 10.25574/cdc.495]{https://doi.org/10.25574/cdc.495}.

\facilities{XRISM(Resolve), CXO, NICER}

\software{Heasoft, XSPEC}

\appendix 

\section{NuSTAR data fits}
NuSTAR spectra corresponding to region 1, with 1' radius, were fit with a single temperature {\tt apec} for the ICM emission and a powerlaw for the AGN emission. The goal of the fitting was to determine the non-thermal contribution from a different instrument with better hard X-ray coverage, so that the reasonableness of XRISM values could be assessed. Table\,\ref{tab:nustarfit} shows the best-fit parameters to region 1. The values for $\Gamma$ and AGN flux are broadly in agreement with Chandra findings, which were ultimately used in the XRISM fit.

\begin{table*}
\centering
\begin{tabular}{lcccccccc}
\hline
Energy& kT &  Fe  &  \multicolumn{2} {c} {ICM Flux}& $\Gamma$ &\multicolumn{2} {c} {AGN Flux}&Cstat/d.o.f.\\
(keV)& (keV) &  (Solar) &  (Full band) & (4-6 keV) &  & (Full band) &  (3-7 keV) &\\
\hline
\\ [-0.95em]
3.0$-$15.0& 2.047$^{+0.135}_{-0.124}$ & 0.817$^{+0.314}_{-0.148}$ & 4.815$^{+0.718}_{-0.508}$  & 1.457$^{+0.058}_{-0.057}$ & 2.09$^{+0.49}_{-0.60}$ & 2.385$^{+0.520}_{-0.732}$    &1.064$^{+0.135}_{-0.134}$   &475.13/501\\ 
\\ [-0.95em]
3.0$-$15.0& 1.996$^{+0.073}_{-0.070}$ & 0.980$^{+0.137}_{-0.113}$ & 3.837$^{+0.120}_{-0.120}$ & 0.985$^{+0.189}_{-0.189}$ & 2.35\footnote{Fixed to Chandra value.} & 2.385$^{+0.133}_{-0.132}$ &1.608$^{+0.152}_{-0.151}$ &475.74/504\\ 
\\ [-0.95em]
3.0$-$11.0& 2.143$^{+0.144}_{-0.112}$ & 0.988$^{+0.437}_{-0.264}$ & 4.220$^{+0.728}_{-0.632}$ & 0.858$^{+0.305}_{-0.303}$  & 2.68$^{+0.32}_{-0.54}$ & 2.698$^{+0.640}_{-0.733}$   &1.725$^{+0.172}_{-0.172}$ &370.85/382 \\ 
\\ [-0.95em]
3.0$-$11.0& 2.165$^{+0.100}_{-0.095}$ & 0.764$^{+0.109}_{-0.092}$ & 4.277$^{+0.201}_{-0.183}$ & 1.185$^{+0.226}_{-0.225}$ & 2.35\footnote{Fixed to Chandra value.} & 1.707$^{+0.191}_{-0.210}$  &1.242$^{+0.175}_{-0.174}$ &370.31/385\\ 
\\ [-0.95em]
\hline
\end{tabular}
\caption{Parameter values obtained from NuSTAR data spectral fitting using single temperature {\tt apec} model plus {\tt powerlaw}. The redshift is fixed to $z=0.00428$ and N$_{H}$ to 1.26 $\times$ 10$^{20}$~cm$^{-2}$. Fluxes are unabsorbed and are given in units of 10$^{-12}$~erg/cm$^2$/s.}\label{tab:nustarfit}
\end{table*}

\section{Density profile}
\begin{figure}
    \centering
    \includegraphics[width=0.5\linewidth]{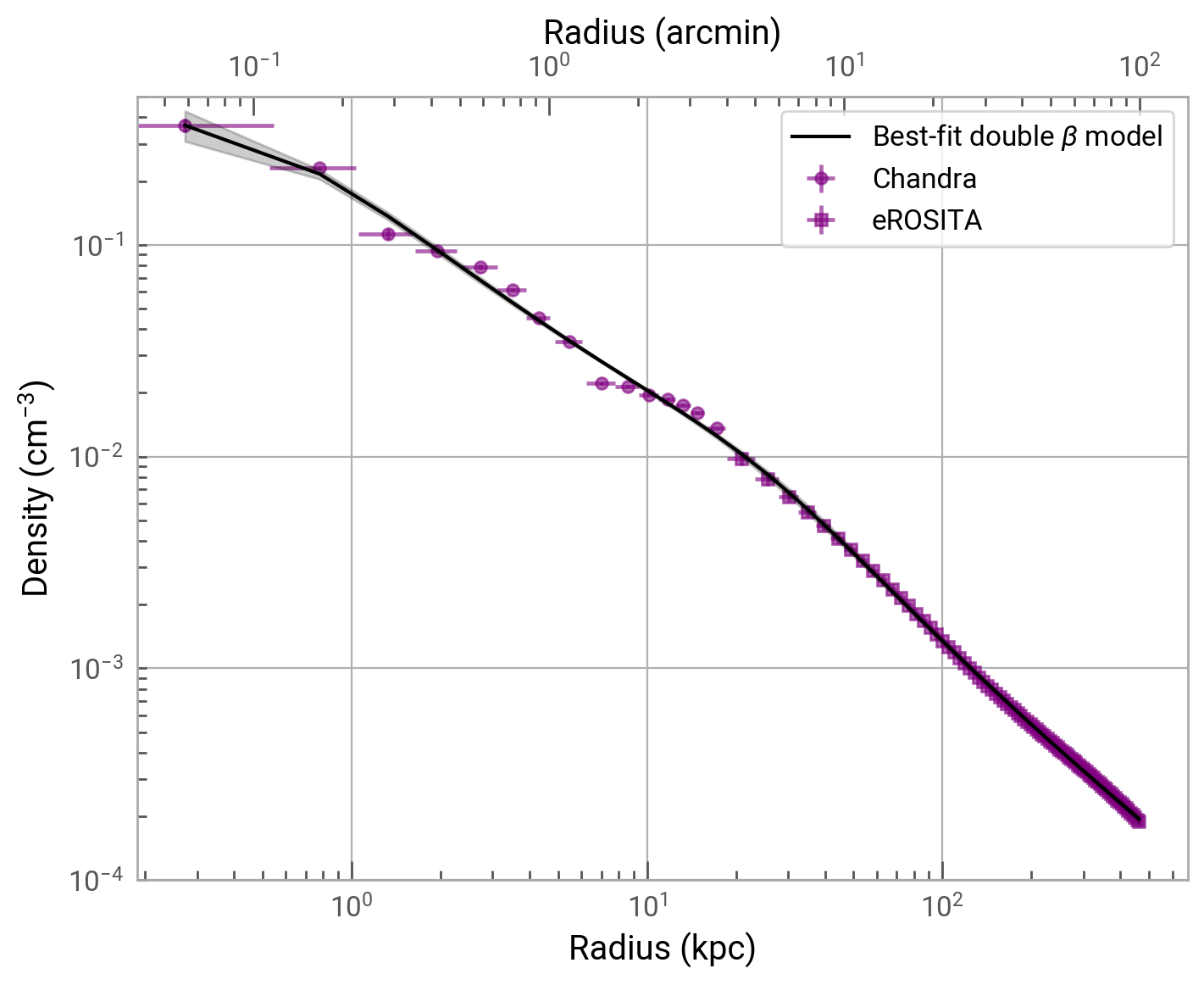}
    \caption{Deprojected density profile and model used for the calculation of effective length. The purple data points up to 20 kpc are from Chandra, while data points beyond 20 kpc are from eROSITA (McCall et. al 2024). The best-fit model (black) is shown with 3$\sigma$ uncertainties for better visibility.}
    \label{fig:density}
\end{figure}

The deprojected density profile is used to calculate cooling rate and effective length $l_{\rm eff}$. Chandra data up to 20\,kpc was combined with eROSITA data to create a profile accounting for cluster emission out to large radii. The profile was then approximated with a double $\beta$-model, which was used for the calculations. The profile and best-fit model are shown in Fig.~\ref{fig:density}.

\bibliographystyle{aasjournalv7}

\begin{thebibliography}{}

  
  

\bibitem[Arevalo et al.(2016)]{Arevalo2016}
  Arevalo, P., Churazov, E., et al.\ 2016, ApJ, On the Nature of X-ray Surface Brightness Fluctuations in M87, 818(1)
  
\bibitem[Arnaud(1996)]{Arnaud1996}
  Arnaud, K. A.\ 1996, in Astronomical Data Analysis Software and
Systems V, 101, 17 

\bibitem[Beaumont et al.(2024)]{Beaumont2024}
  Beaumont, S., et al.\ 2024, A\&A, 686, A41

\bibitem[Belsole et al.(2001)]{Belsole2001}
  Belsole, E., et al.\ 2001, A\&A, 365, L188

\bibitem[B\"ohringer et al.(1994)]{Boehringer1994}
  B\"ohringer, H., et al.\ 1994, Nature, 368, 828

\bibitem[B\"ohringer et al.(2001)]{Boehringer2001}
  B\"ohringer, H., et al.\ 2001, A\&A, 365, 181

\bibitem[Boselli et al.(2019)]{Boselli2019}
  Boselli, A., et al.\ 2019, A\&A, 623, A52. doi:10.1051/0004-6361/201834492

\bibitem[Br\"uggen \& Vazza(2015)]{Brueggen2015}
  Brüggen, M., \& Vazza, F.\ 2015, Magnetic Fields in Diffuse Media, 407, 599

\bibitem[Cappellari et al.(2011)]{Cappellari2011}
  Cappellari M., et al.\ 2011, MNRAS, 413, 813

\bibitem[Cash(1979)]{Cash1979}
  Cash W.\ 1979, ApJ, 228, 939

\bibitem[Chatzigiannakis et al.(2025)]{Chatzigiannakis2025}
  Chatzigiannakis, D., et al.\ 2025, submitted

\bibitem[Churazov et al.(2001)]{Churazov2001}
  Churazov, E., et al.\ 2001, ApJ, 554, 261

  \bibitem[Churazov et al.(2008)]{Churazov2008}
  Churazov, E., et al.\ 2008, MNRAS, 388, 1062

\bibitem[Churazov et al.(2010)]{Churazov2010}
  Churazov, E., Zhuravleva, I., Sazonov, S., Sunyaev, R.\ 2010, Space\ Sci.\ Rev., 157, 193

  \bibitem[Cucchetti et al.(2019)]{Cucchetti2019}
  Cucchetti, E., et al.\ 2019, A\&A, 629, A144

\bibitem[Fabian et al.(2017)]{Fabian2017} Fabian A.~C., et al.\ 2017, MNRAS, 464, L1

\bibitem[Ford \& Butcher(1979)]{Ford1979}
  Ford, H.C. and Butcher, H. 1979, ApJS, 41, 147

\bibitem[Forman et al.(2005)]{Forman2005}
  Forman, W., et al.\ 2005, ApJ, 635, 894

\bibitem[Forman et al.(2007)]{Forman2007}
  Forman, W., et al.\ 2007, ApJ, 665, 1057

  \bibitem[Forman et al.(2017)]{Forman2017}
  Forman, W., et al.\ 2017, ApJ, 844, 122

\bibitem[Freeman et al.(2002)]{Freeman2002}
  Freeman, P.E., Kashyap, V., Rosner, R., Lamb, D.Q., 2002, ApJS, 138, 185

\bibitem[Fruscione et al.(2006)]{Fruscione2006}
  Fruscione, A., et al.\ 2006, in Society of Photo-Optical Instrumentation Engineeres (SPIE) Conference Series, doi:10.1117/12.671760

\bibitem[Gaspari et al.(2013)]{Gaspari2013a}
  Gaspari, M., et al.\ 2013, MNRAS, 432, 4, 3401, doi:10.1093/mnras/stt692

\bibitem[Gatuzz et al.(2022)]{Gatuzz2022}
  Gatuzz, E., et al.\ 2022, MNRAS, 511(3), 4511

\bibitem[Gilfanov et al.(1986)]{Gilfanov1986}
  Gilfanov, M.R., et al.\ 1986, Soviet Astronomy Letters, 13, 3
    
\bibitem[HI4PI et al.(2016)]{HI4PI}
  HI4PI Collaboration\ 2016, A\&A, 594, 116

\bibitem[Harris et al.(2003)]{Harris2003}
  Harris D. E., Biretta, J.A., et al.\ 2003, ApJ, Flaring X-Ray Emission from HST-1, a Knot in the M87 Jet, 586(1), L41
  
\bibitem[Harris et al.(2009)]{Harris2009}
  Harris D. E., Cheung C. C., Stawarz \L., Biretta J. A., Perlman
  E. S.\ 2009, ApJ, 699, 305

\bibitem[Harris et al.(2006)]{Harris2006}
  Harris, D. E., et al.\ 2006, ApJ, 640, 211

\bibitem[Harrison et al.(2013)]{Harrison2013}
  Harrison, F. A., et al.\ 2013, ApJ, 770(2), 103

\bibitem[Heinrich et al.(2024)]{Heinrich2024}
  Heinrich, A., et al.\ 2024, MNRAS, 528(4), 7274–7299

\bibitem[Hitomi Collaboration et al.(2016)]{Hitomi2016} Hitomi Collaboration, Aharonian, F., Akamatsu, H., et al.\ 2016, \nat, The quiescent intracluster medium in the core of the Perseus cluster, 535, 7610, 117. doi:10.1038/nature18627

\bibitem[Irwin et al.(2003)]{Irwin2003}
  Irwin, J.A., Athey, A.E., Bregman, J.N.\ 2003, ApJ, 587, 356

\bibitem[Kaastra et al.(1996)]{Kaastra1996}
  Kaastra, J. S., \& Mewe, R., \& Nieuwenhuijzen, H. \ 1996, UV and X-ray Spectroscopy of Astrophysical and Laboratory Plasmas,
  ed. K. Yamashita \& T. Watanabevol,\ 411–414  

\bibitem[Kaastra et al.(2016)]{Kaastra2016}
  Kaastra, J.S., Bleeker, J.A.M.\ 2016, A\&A, 587, A151

\bibitem[Kalberla et al.(2005)]{Kalberla2005}
  Kalberla, P. M., et al.\ 2005, A\&A, 440, 775

\bibitem[Kaneda et al.(2003)]{Kaneda2003}
  Kaneda, Y., et al.\ 2003 Phys. Fluids, 15, L21

\bibitem[Kilbourne et al.(2018)]{Kilbourne2018}
  Kilbourne, C.A., et al.\ 2018, PASJ, 70(2), 18
  
\bibitem[Kolmogorov(1941)]{Kolmogorov1941}
  Kolmogorov, A. 1941 Dokl. Akad. Nauk SSSR, 30, 301

\bibitem[Li et al.(2020)]{Li2020}
  Li, C., et al.\ 2020, MNRAS, 492, 2, 2775. doi:10.1093/mnras/staa027

\bibitem[Lodders et al.(2009)]{Lodders2009}
  Lodders, K., Palme, H., \& Gail, H.P.\ 2009, in The Solar System ed.\ J.E. Tr\"umper (Springer-Verlag Berlin Heidelberg) 4B, 712

\bibitem[Markevitch et al.(2007)]{Markevitch2007}
  Markevitch, M., Vikhlinin, 2007

\bibitem[McCall et al.(2024)]{McCall2024}
  McCall, H., et al.\ 2024, A\&A, 689, A113

\bibitem[McNamara et al.(2016)]{McNamara2016}
  McNamara, B.R., et al.\ 2016, ApJ, 830, 2, 79. doi:10.3847/0004-637X/830/2/79

\bibitem[Mei et al.(2007)]{Mei2007}
  Mei, S., et al.\ 2007, ApJ, 655, 144

\bibitem[Molin et al.(2025)]{Molin2025}
  Molin, A., et al.\ 2025, A\&A, submitted, arXiv:2505.14378

\bibitem[Pinto et al.(2015)]{Pinto2015}
  Pinto, C., et al.\ 2015, A\&A, 575, A38

\bibitem[Porter et al.(2024)]{Porter2024}
  Porter, F.S., Kilbourne, C.A., Chiao, M., et al.\ 2024, SPIE, 13093, 130931

\bibitem[Revnivtsev et al.(2014)]{Revnivtsev2014}
  Revnivtsev, M.G., Sunyaev, R.A., Krivonos, R.A., Tsygankov, S.S., Molkov, S.V.\ 2014, Astronomy Letters, 40, 22

\bibitem[RojasBolivar et al.(2023)]{Rojas23}
  Rojas Bolivar, et al. \ 2015, ApJ, 954, 76

\bibitem[Rose et al.(2025)]{Rose2025}
    Rose, T., McNamara, B.R., et al.\ 2025, ApJ, submitted, arXiv:2505.01494
  
\bibitem[Russell et al.(2015)]{Russell2015}
  Russell, H. R., Fabian, A.C., McNamara, B.R., Broderic, A.E.\ 2015, MNRAS, 451(1), 588

\bibitem[Russell et al.(2018)]{Russell2018}
  Russell, H. R., et al.\ 2018, MNRAS, 477, 3583

\bibitem[Sarzi et al.(2018)]{Sarzi2018}
  Sarzi, M., et al.\ 2018, MNRAS, 478, 3, 4084. doi:10.1093/mnras/sty1092

\bibitem[Sanders et al.(2020)]{Sanders2020}
  Sanders, J.S., et al.\ 2020, A\&A, 633, A42

\bibitem[Simionescu et al.(2007)]{Simionescu2007}
  Simionescu, A., et al.\ 2007, A\&A, 465, 749

\bibitem[Simionescu et al.(2017)]{Simionescu2017}
  Simionescu, A., et al.\ 2017, MNRAS, 469(2), 1476

\bibitem[Simionescu et al.(2018)]{Simionescu2018}
  Simionescu, A., et al.\ 2018, MNRAS, 475(3), 3004. doi:10.1093/mnras/sty047


\bibitem[Simionescu et al.(2019)]{Simionescu2019}
  Simionescu, A., et al.\ 2019, Space\ Sci.\ Rev., 215, 24

\bibitem[Sparks et al.(1993)]{Sparks1993}
  Sparks, W.B., Ford, H.C., and Kinney, A.L. 1993, ApJ, 413, 531

\bibitem[Sparks et al.(1996)]{Sparks1996}
  Sparks, W.B., Biretta, J.A., and Macchetto, F.\ 1996, ApJ, 473, 254

\bibitem[Sparks et al.(2004)]{Sparks2004}
  Sparks, W.B., Donahue, M., Jord{\'a}n, A. et al.\ 2004, ApJ, 607(1), 294

\bibitem[Sparks et al.(2009)]{Sparks2009}
  Sparks, W.B. et al.\ 2009, ApJL, 704(1), L20

\bibitem[Sparks et al.(2012)]{Sparks2012}
  Sparks, W.B. et al.\ 2012, ApJL, 750(1), L5
  
\bibitem[Sreenivasan et al.(1995)]{Sreenivasan1995}  
  Sreenivasan, K. R. 1995, Phys. Fluids, 7, 2778

\bibitem[Stofanova et al.(2021)]{Stofanova2021}
  \v{S}tofanov\'{a}, L., et al.\ 2021, \aap, 655, 2

\bibitem[Urban et al.(2011)]{Urban2011}
  Urban, O., et al.\ 2011, MNRAS, 414, 2101

\bibitem[Voit et al.(2015)]{Voit2015}
  Voit, G.M., et al.\ 2015, Nature, 519, 7542. doi:10.1038/nature14167

\bibitem[Wachter et al.(1979)]{Wachter1979}
  Wachter K., Leach R., Kellogg E.\ 1979, ApJ, 230, 274

\bibitem[Werner et al.(2013)]{Werner2013}
  Werner N., et al.\ 2013, ApJ, 767, 153. doi:10.1088/0004-637X/767/2/153

\bibitem[Werner et al.(2016)]{Werner2016}
  Werner N., et al.\ 2016, MNRAS, 455, 846

\bibitem[Wik et al.(2014)]{Wik2014}
  Wik, D. R., et al.\ 2014, ApJ, 792(1), 48

\bibitem[XRISM Collaboration et al.(2024)]{XRISM2024}
  XRISM Collaboration\ 2024, PASJ, 76(6), 1186-1201

\bibitem[XRISM Collaboration et al.(2025)]{XRISM2025a} XRISM Collaboration, Audard, M., Awaki, H., et al.\ 2025, \apjl, XRISM Reveals Low Nonthermal Pressure in the Core of the Hot, Relaxed Galaxy Cluster A2029, 982, 1, L5. doi:10.3847/2041-8213/ada7cd

\bibitem[XRISM Collaboration et al.(2025)]{XRISM2025b} XRISM Collaboration, Audard, M., Awaki, H., et al.\ 2025, \apjl, XRISM Forecast for the Coma Cluster: Stormy, with a Steep Power Spectrum, 985, 1, L20. doi:10.3847/2041-8213/add2f6

\bibitem[XRISM Collaboration et al.(2025)]{XRISM2025c} XRISM Collaboration, Audard, M., Awaki, H., et al.\ 2025, \nat, The bulk motion of gas in the core of the Centaurus galaxy cluster, 638, 8050, 365. doi:10.1038/s41586-024-08561-z

\bibitem[XRISM Collaboration et al.(2025)]{XRISM2025d}
  XRISM Collaboration, Audard, M., Awaki, H., et al.\ 2025, PASJ, Constraining gas motion and non-thermal pressure beyond the core of the Abell 2029 galaxy cluster with XRISM, accepted. 

\bibitem[XRISM Collaboration et al.(2025)]{XRISM2025e}
  XRISM Collaboration, Audard, M., Awaki, H., et al.\ 2025, Nature, Disentangling multiple gas kinematic drivers in the Perseus Galaxy Cluster, submitted. 

\bibitem[Young et al.(2002)]{Young2002}
  Young, A. J., Wilson, A. S., and Mundell, C. G. 2002, ApJ, 579, 560

\bibitem[Zhuravleva et al.(2014)]{Zhuravleva2014}
  Zhuravleva, I., et al.\ 2014, Nature, 515, 85

  \bibitem[Zhuravleva et al.(2012)]{Zhuravleva2012}
  Zhuravleva, I., et al.\ 2012, MNRAS, 422, 2712


\bibitem[ZuHone et al.(2013)]{ZuHone2013}
  ZuHone, J. A., et al.\ 2013, ApJ, 762, 78

\bibitem[ZuHone et al.(2015)]{ZuHone2015}
  ZuHone, J. A., et al.\ 2015, ApJ, 798, 90

  
\end{thebibliography}

\allauthors
\end{document}